\documentclass[11pt,a4paper,doubleblind]{article}

\usepackage[a4paper, left=1in, right=1in, top=1in, bottom=1in, includehead, includefoot, foot = 0pt, head=0pt]{geometry} 
\usepackage[group-separator={,},group-digits=integer]{siunitx} 
\usepackage{booktabs} 
\usepackage{enumitem} 

\usepackage[round,colon,authoryear]{natbib} 
\usepackage[pdftex,colorlinks,citecolor=blue,urlcolor=blue,linkcolor=red]{hyperref} 
\usepackage{graphicx} 
\parskip=\medskipamount 
\usepackage[small,bf,hang]{caption}	
\usepackage{amssymb} 
\usepackage{amsmath} 
\usepackage{bbm} 
\usepackage{amsbsy} 
\usepackage{mathrsfs} 
\usepackage[linesnumbered,ruled,vlined]{algorithm2e} 
\SetAlCapFnt{\small} 
\usepackage{multirow} 
\usepackage[dvipsnames]{xcolor} 
\usepackage{subcaption} 
\usepackage[hang,flushmargin]{footmisc} 
\usepackage{tikz} 
\usetikzlibrary{matrix} 
\usepackage{rotfloat}
\usepackage{comment}
\usepackage{multirow}
\usepackage{setspace}
\usepackage{eurosym}
\usepackage[normalem]{ulem}

\usepackage{authblk}

\usepackage{fancyhdr}
\graphicspath{{Figures/}}

\begin{document}

\title{\LARGE  Social network analytics for supervised fraud detection in insurance}
\author[1]{Mar\'{i}a \'{O}skarsd\'{o}ttir}
\author[2] {Waqas Ahmed}
\author[3,4,5,6]{Katrien Antonio}
\author[3,7]{Bart Baesens}
\author[2]{R\'{e}mi Dendievel}
\author[2]{Tom Donas}
\author[3,5]{Tom Reynkens}

\affil[1]{Department of Computer Science, Reykjavik University, Iceland.}
\affil[2]{AG Insurance, Belgium.}
\affil[3]{Faculty of Economics and Business, KU Leuven, Belgium.}
\affil[4]{Faculty of Economics and Business, University of Amsterdam, The Netherlands.}
\affil[5]{LRisk, Leuven Research Center on Insurance and Financial Risk Analysis, KU Leuven, Belgium.}
\affil[6]{LStat, Leuven Statistics Research Center, KU Leuven, Belgium.}
\affil[7]{Dept. of Decision Analytics and Risk, University of Southampton, United Kingdom.}
\date{}
\maketitle
 \thispagestyle{empty}
\begin{abstract}
\noindent
Insurance fraud occurs when policyholders file claims that are exaggerated or based on intentional damages.
This contribution develops a fraud detection strategy by extracting insightful information from the social network of a claim. First, we construct a network by linking claims with all their involved parties, including the policyholders, brokers, experts, and garages.
Next, we establish fraud as a social phenomenon in the network and use the BiRank algorithm with a fraud specific query vector to compute a fraud score for each claim. From the network, we extract features related to the fraud scores as well as the claims' neighborhood structure.
Finally, we combine these network features with the claim-specific features and build a supervised model with fraud in motor insurance as the target variable.
Although we build a model for only motor insurance, the network includes claims from all available lines of business. Our results show that models with features derived from the network perform well when detecting fraud and even outperform the models using only the classical claim-specific features. Combining network and claim-specific features further improves the performance of supervised learning models to detect fraud.
The resulting model flags highly suspicions claims that need to be further investigated. Our approach provides a guided and intelligent selection of claims and contributes to a more effective fraud investigation process.
\end{abstract}

\section{Introduction}\label{sec:introduction}
This paper presents a big-data risk analytics approach to detect suspicious insurance claims with an analytical model that combines the information from a network of insurance claims and involved parties with the traditional claim-specific features collected by the insurer.
Insurance companies bring security to society by offering protection against financial losses. By organizing and
managing pools of risks they enable individuals and companies to transfer risk in exchange for a premium.
Fraud, the wrongful or criminal deception intended for financial or personal gain, is a considerable concern in many industries, such as in telecommunication and healthcare.  Other types and areas of fraud include corruption, money laundering, tax evasion, identity theft and credit card fraud, to name a few \citep{baesens2015fraud}. The insurance industry is no exception. Insurance fraud is encountered when policyholders file claims that are exaggerated, fake or based on intentional damage with the intention to receive compensation from the insurance company.
In non-life insurance, the total yearly cost of fraudulent claims is estimated at over \euro1 billion in the UK and \$30 billion in the United States, which, in the latter case, corresponds to an increased yearly premium of between \$200 and \$300 for an average family \citep{fbi, australia}.
In Australia, fraud has been identified as the most costly crime category, with insurance fraud amounting a total of AU\$2.1 billion every year, equivalent to AU\$73 per insurance policy \citep{australia}.
Furthermore, the Dutch Association of Insurers reports a saving of \euro 79 million in 2015 and 82 million in 2018 because of fraud identification efforts \citep{vvv,vvv2019}.
Undeniably, it is in the interest of both insurance companies and their customers to detect fraudulent activity to keep insurance contracts affordable.

When an insurance claim is flagged because of suspicion of fraud, fraud investigators use their expertise to confirm whether the claim is fraudulent or not \citep{warren2018lying}.
The insights obtained from such investigations are then used to help detect other suspicious claims.
This comprises the fraud detection cycle. 
The rare occurrence of fraud on the one hand and the many uninvestigated cases or claims on the other hand, are important challenges in the construction of fraud detection methods \citep{jensen1997prospective,baesens2015fraud}. 
The uncommon nature of fraud creates imbalanced data sets, where fraudulent claims are at risk of being treated as insignificant in analytical models, causing a bias towards the non-fraudulent claims. 
In addition, organizations have typically neither the time nor the resources to investigate all (suspicious) claims and to label each claim as fraudulent or legitimate \citep{warren2018lying}. As a result, only a small fraction of all claims have a known label. 
A fraud detection model is an efficient way to lead the investigators in the direction of fraudulent claims. By using techniques such as business rules or analytical models, only the most suspicious claims can be given to the investigator for further inspection. This avoids spending precious time on investigating non-fraudulent claims.
Insurance fraud materializes in various ways; garages and hospitals may for example overcharge for material and work. Here, we consider fraud committed by policyholders when filing claims, such as false declarations, exaggeration of damage and intentional damage. 

In the insurance industry, fraud detection models are based on business rules as well as analytical techniques \citep{EIOPA}.
Business rules are usually formed by intuition, past experiences and input from the experts that conduct the investigations.
Both unsupervised and supervised machine learning methods have been successfully used in the literature on insurance fraud detection strategies.
A first and prominent fraud detection technique is PRIDIT, which computes a claim-specific fraud suspicion score via principal component analysis of so-called RIDIT scores \citep{brockett2002fraud,ai2009assessing}. 
RIDIT transforms a claim's available categorical features into a set of numerical features, reflecting the relative abnormality of the observed categorical features. As such, the scores represent the likelihood of fraud and can be used as input in further statistical analysis.
\cite{stripling2018isolation,ghorbani2018fraud,nian2016auto,hainaut2018self} use principles from anomaly detection for unsupervised fraud detection strategies that aim at finding unusual and uncommon behavior deviating from the norm in the dataset.
Approaching insurance fraud detection as a classification problem, \cite{viaene2002comparison} present a benchmark study of classification techniques.
This supervised learning perspective has the advantage of learning fraud patterns in the data but requires past knowledge about fraudulent cases, which is rare because of the high class imbalance. Several fraud detection approaches tackle the problem of class imbalance by rebalancing the dataset using sampling techniques, such as under- and oversampling, SMOTE or ROSE \citep{subudhi2017use, van2016gotcha,sundarkumar2015novel}.
In a semi-supervised approach, \cite{botelho2011combining} use unlabeled data to reinforce the learning algorithm.
Other researchers strive at optimally adapting known techniques to the fraud detection problem \citep{li2018principle} or enriching the data with alternative information.
\cite{kose2015interactive} use machine learning methods to incorporate expert knowledge in the fraud detection model to make the learning of fraud patterns more interactive.
\cite{wang2018leveraging} mine experts' text descriptions of accidents and apply deep learning to detect fraud in car insurance whereas \cite{vsubelj2011expert} use networks to detect groups of collaborating fraudsters.

Fraudulent activity can often be associated with organized schemes carried out by collaborating fraudsters that go to great lengths in order to hide their tracks while maximizing their gain \citep{van2016gotcha}.
In criminology, it is well known that two thirds of criminals commit crimes together \citep{reiss1988co} in order to maximize their reward and to minimize the risk of getting caught \citep{andresen2009impact}.
In particular, in organized crime, such as fraud, social relations are of a great importance as they not only provide access to co-offenders and profitable opportunities but also invigorate the necessary trust \citep{van2009criminal}.
Police departments are therefore examining criminal networks to a greater extent to detect organized crime groups and their activity with social network analytics \citep{tayebi2016social}.
Social network analytics is the analysis of social structures, i.e. the ties between individuals and organisations, using networks and graph theory \citep{newman2010networks}.
In fraud detection, this activity can be detected for example by measuring proximity to known fraudulent cases or exposure to fraud influence \citep{jamshidi2012efficient}.
In a first attempt of using this approach for fraud detection in insurance, \cite{vsubelj2011expert} argue that groups of collaborating fraudsters can only be detected with the appropriate representation, i.e. social networks, since they are often related to each other in some way.
They use collision data from police records to build networks, rely on expert knowledge to identify suspicious sub-networks using PRIDIT scores and propose an iterative assessment algorithm to detect the most suspicious entities in each sub-network.

This paper addresses risk management in the context of insurance fraud.
We propose a framework to build and to evaluate an analytical fraud detection model for insurance claims.
We hereby combine the socio-demographic information about the policyholder and classic information about the claim with information about the claim's social network structure.
The social network connects a claim with all its involved parties, e.g., policyholders, brokers, garages and experts, which in turn are connected to all claims they are involved in.
In this way, we build a bipartite network with a few million insurance claims and parties that is a fitting representation of the relationships between the various parties by means of the claims in which they are involved.
As such, we are able to look beyond the classical properties of the claim, the policyholder and the policy, and study the social structures of collaborating fraudsters in insurance fraud detection tools and models.
The network contains claims from several types of insurance covers, e.g., motor, fire, and liability, which provides a holistic view on the customers of the insurance company. In the actual fraud detection model, we focus on the motor line of business.

We start by establishing fraud as a social phenomenon in this network and show that fraudulent claims are connected to other fraudulent claims to a greater extent.
The collaboration of fraudsters in fraud circles to hide their tracks and lower the risk of getting caught, is a fundamental assumption for a network based fraud detection approach, and is known as homophily in social sciences.
Homophily refers to people's tendency to associate with others whom they perceive as being similar to themselves in some way.
In a network of claims, this means that the fraudulent claims are more often connected to other fraudulent claims and that non-fraudulent claims are more likely to be connected to other non-fraudulent claims.
In addition, fraudulent and non-fraudulent claims are connected to each other to a lesser degree.

Having empirical evidence of homophily, we rank the claims in the network relative to known fraudulent claims.
In essence, the goal of node ranking is to assign a score to each node in accordance with a specific criterion.
This became a prominent research topic as internet search engines were emerging, starting with PageRank and HITS \citep{kleinberg1999authoritative, page1999pagerank}. 
Applications of node ranking are omnipresent including social analysis of Twitter accounts \citep{kwak2010twitter}, connectivity patterns in neuroscience \citep{lohmann2010eigenvector} as well as fraud detection in online lending, tax evasion and social security \citep{min2018behavior,botelho2011combining,van2016gotcha}.
We use the BiRank algorithm proposed by \cite{he2017birank}  as an extension of the PageRank algorithm tailor made for bipartite networks.
In addition, inspired by \cite{van2016gotcha}, we modify the BiRank algorithm by adapting the query vector to include the knowledge of known fraudulent claims in the network.
This steers the ranking towards the fraud in the network and results in higher scores for claims that are closer and more densely connected to known fraudulent claims.
In this way, all nodes in the network obtain a score which represents proximity to fraudulent claims. We use the scores to compute a set of features which characterize the position of a claim in the network, such as, the average score of all parties involved in the claim.
We combine the scores and the features from the network with the traditional features of the claim, the policyholder and the policy in an analytical fraud detection model to flag highly suspicions claims that need to be investigated further.
The approach provides a guided and intelligent selection of claims and thus contributes to a more effective fraud investigation process.
As such, we demonstrate how to harness high-volume, high-dimensional and multi-source data through network science.
Using these data in a predictive machine learning model allows to unravel their potential to detect and to prevent insurance fraud via big-data analytics.

The rest of this paper is organized as follows.
In the next section, we present the dataset used in this study and demonstrate how to convert it into a bipartite network.
Moreover, we discuss fraudulent structures and show empirical evidence of homophily among fraudulent claims.
In Section~\ref{sec:frauddetectionmodel}, we elaborate on our fraud detection model and in Section~\ref{sec:experimentalsetup} we explain the setup of the subsequent experiments.
In Section~\ref{sec:results}, we present the results of our proposed technique and finally, conclude our research in Section~\ref{sec:conclusion}.


\section{Insurance fraud dataset}\label{sec:dataset}
\subsection{Description of the data}\label{sec:localfeatures}
We analyze a dataset with a couple of million claims filed by an insurance company's policyholders over a period of a few years.
Claims are formal requests sent to an insurance company for coverage or compensation of a covered loss.
The dataset includes information on the insured event, such as the claim amount, as well as information on the policyholder and their claims history.
Our goal is to build a fraud detection model for the motor insurance cover.
Therefore, Table \ref{T:dataset} details the available features for the motor insurance claims and their meaning. 
We refer to these features as \emph{intrinsic} features.
They are specific to the line of business (LoB) for which the fraud detection model is designed.

In addition to the intrinsic features, each claim has a list of the people and companies involved in the claim.
These are policyholders, brokers, experts and garages.
Policyholders are people or companies that are involved in the incident in one of the following ways.
A policyholder may be covered by the insurer and request compensation using their own policy, or the policy of another client of the insurance company.
Alternatively, they may be involved in a claim (e.g.,~as a victim) while not being covered by the insurance company.
Insurance brokers are individuals or companies that sell, solicit and negotiate insurance on behalf of policyholders.
Experts evaluate the loss filed via an insurance claim and garages repair damaged cars.
In this paper the term \emph{party} refers to all these entities that are not claims.
We use claims and involved parties to build a network, connecting claims and parties across all available LoBs.

\begin{table}
\centering
\caption{Description of the available intrinsic features for the motor insurance claims.}\label{T:dataset}
\scalebox{0.8}{
\begin{tabular}{p{2.4cm}lp{13cm}}
\toprule
Type&Feature&Description\\
\midrule
Target&  \texttt{fraud}&Is the claim fraudulent: yes, no or unknown.\\ \midrule
\multirow{24}{2cm}{Policyholder characteristics}&
  \texttt{age}&Age of policyholder when incident occurred.\\
&  \texttt{responsibilityCode}&Policyholder's responsibility in the incident: at fault, shared responsibility or full right.\\
&  \texttt{numContracts}&Number of contracts the policyholder has or has had with the insurer.\\
&  \texttt{claimAge}& Number of months from beginning of contract to the date the incident occurred. \\
&  \texttt{nClaims1}&Number of claims across all LoBs in the last year before the incident occurred.\\
&  \texttt{nClaims5}&Number of claims across all LoBs in the last 5 years before incident occurred.\\
&  \texttt{lastClaim}&Number of months since the last claim occurrence.\\
&  \texttt{amount1}&Claimed amount across all LoBs in the last year before current claim occurrence.\\
&  \texttt{amount5}&Claimed amount across all LoBs in the last 5 years before current claim occurrence.\\
&\texttt{refused1}&Number of times compensation was refused in the last year before current claim occurrence.\\
&  \texttt{refused5}&Number of times compensation was refused in the last 5 years before current claim occurrence.\\
&  \texttt{atfault1}&Number of times the policyholder had responsibility code `at fault' in the last year before current claim occurrence.\\
&  \texttt{atfault5}&Number of times the policyholder had responsibility code `at fault' in the last 5 years before current claim occurrence.\\
&  \texttt{samesits1}&Number of times the policyholder had the same responsibility code as the current claim in the last year before current claim occurrence.\\
&  \texttt{samesits5}&Number of times the policyholder had the same responsibility code as the current claim in the last 5 years before current claim occurrence.\\ \midrule
\multirow{6}{2cm}{Claim characteristics}&
  \texttt{claimAge}& Number of months from beginning of contract to the date the incident occurred. \\
&  \texttt{people}&Number of people involved in the claim.\\
&  \texttt{company}&Number of companies involved in the claim.\\
&  \texttt{police}&Police was called when the incident happened: yes or no.\\
&  \texttt{daysReport}&Number of days from the occurrence of incident to the filing of the claim.\\
&  \texttt{amount}&The claimed amount for closed claims or the expected claimed amount if the claim is still open.\\
\bottomrule
\end{tabular}}
\end{table}

To establish whether a claim is fraudulent or not, it must undergo an investigation by fraud inspectors at the insurance company.
These inspectors label the claims as fraudulent or non-fraudulent after an inquiry into the claim.
However, due to time constraints and lack of resources, only a small fraction of all claims undergo such an investigation.
This means that most claims are settled without an investigation. Hence, a number of fraudulent claims will remain undetected.
We refer to claims that have been investigated as known, having a known label or as being labeled, whereas claims that were never investigated are referred to as unlabelled, unknown, or, having an unknown label.
Per year in our dataset, less than a percent of all claims go through a formal fraud investigation with less than half resulting in a known fraud label.

\subsection{A bipartite network of insurance claims and involved parties}\label{sec:bipartNetwork}
\begin{figure}
\centering
\includegraphics[scale=1]{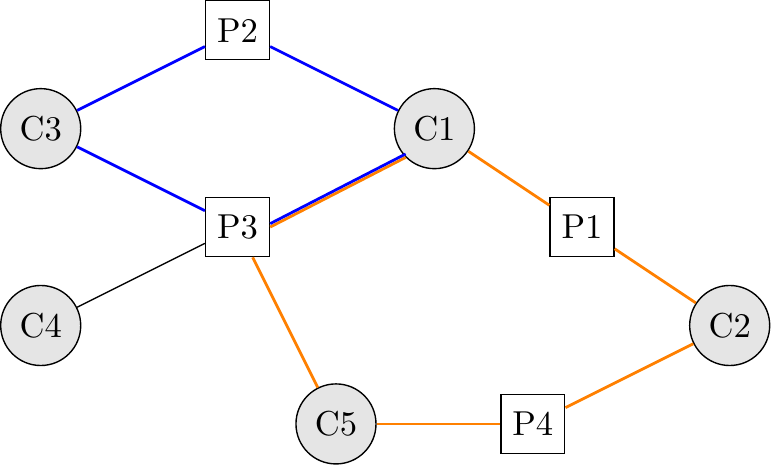}
\caption{A sample network of claims and parties. The circles are claims and the rectangles are parties. The nodes connected with blue edges form a 4-cycle and the nodes connected with orange edges form a 6-cycle. \label{simpleNetwork} }
\end{figure}

We use the claims and the involved parties to build a bipartite network, including both open and closed claims reported within a certain period.
When a claim is filed, we link to it all the involved parties.
The involved parties in a motor insurance claim could be, e.g., the policyholder who owns the car, the broker who solicits the claim and the garage where the car is repaired.
Furthermore, we link each party to all the claims in which this party is involved.
We refer to the network as a social network.
Figure \ref{simpleNetwork} shows a sample network of five claims and four parties. Claim C1 is linked to its involved parties, P1, P2 and P3. Each party is then also linked to the claims that they are involved in. Party P1 is linked to claim C2, party P2 is linked to claim C3, and party P3 is linked to claims C3, C4 and C5.
This network is not restricted to motor insurance claims only, but includes claims from all lines of business in the non-life sector of the insurance company, i.e., motor, fire, liability and work accidents. As such, the social network represents a holistic view on the riskiness of clients.

The network is a bipartite network because it is composed of two types of nodes, i.e., claims ($C$) and parties ($P$), with no edges between nodes of the same type.
Formally, we let $G=(C\cup P,E)$ be a bipartite network of nodes $C\cup P$ and edges $E$.
Each node belongs to exactly one of the sets $C$ and $P$ and each edge in $E$ connects one node in $C$ to one node in $P$ \citep{newman2010networks}.
The nodes in the set $C$ correspond to insurance claims and the nodes in $P$ are the various parties involved in the claims, that is, the policyholders, brokers, experts and garages.
We let $n_C$ be the number of nodes in $C$ and denote individual nodes, i.e. claims, with $c_i$ where $i\in\{1, \dots, n_C\}$.
Similarly, $n_P$ is the number of nodes in $P$ and individual nodes, i.e. parties, are given by $p_j, j\in\{1, \dots, n_P\}$.
The network's edges carry weights to indicate the strength of the connection between the two nodes they connect.
All the edges are represented collectively by a weight matrix $\boldsymbol{W}=(w_{ij})$, where $i\in\{1, \dots, n_C\}, j\in\{1, \dots, n_P\} $, with $n_C$ rows and $n_P$ columns, corresponding to the nodes in $C$ and $P$ respectively.
Thus, if node $c_i$ is connected to node $p_j$ the value of $w_{ij}$ is positive and zero otherwise.
When the network is unweighted, all non-zero values in $\boldsymbol{W}$ are equal to one.
The network is undirected as no direction is given to the edges, that is $w_{ij}=w_{ji}, \forall i,j \in n_C\cup n_P$.

The $k$-th order neighborhood of a node $c_i$, denoted with $\mathcal{N}_{c_i}^k$, is the set of all nodes that are connected to $c_i$, via a path of exactly $k$ edges.
Thus, the first order neighborhood of a claim $c_i$ is the set of all parties that are involved in the claim, or
\[
\mathcal{N}_{c_i}^1=\{p_j|w_{ij}\ne 0\},
\]
whereas the second order neighborhood of a claim is the set of all claims that are involved in the parties in $\mathcal{N}_{c_i}^1$, or
\[
\mathcal{N}_{c_i}^2=\{c_k|p_j \in \mathcal{N}_{c_k}^1 \wedge w_{kj}\ne 0\}\setminus c_i.
\]

The degree of a node, denoted with $d_i$ for claim $c_i$ or $d_j$ for party $p_j$, is the number of nodes in the node's first order neighborhood for an unweighted network and the sum of weights on the edges between the node and the nodes in its first order neighborhood for a weighted network.
We use the diagonal matrix $\boldsymbol{D_C}$ to denote the degree of all claims, i.e., $(\boldsymbol{D_C})_{ii}=d_i$, and similarly, the diagonal matrix $\boldsymbol{D_P}$ for the degree of all parties. The matrices $\boldsymbol{D_C}$ and $\boldsymbol{D_P}$ are square matrices of orders $n_C$ and $n_P$, respectively.


Table \ref{T:edgeSummary} shows summary statistics for the nodes and edges in the network.
Set $C$ contains only one type of nodes, namely the claims. The average and median of the number of parties involved in a claim is 3.79 and 3, respectively.
There are four types of party nodes, namely, policyholders, brokers, experts and garages.
With a relative frequency of $96.36\%$, policyholders are the most common parties and they are connected to two claims on average.
Broker nodes have a relative frequency of only $0.39\%$ and yet some of them are very important in the network as is shown by their high edge relative frequency.
Out of all the edges, $27.82\%$ are connected to a broker and the broker degree ranges from one to almost twenty thousand.
Experts show a similar trend in terms of node relative frequency, but their edge relative frequency is lower.
Finally, although there is a higher number of garages, their connectivity is less than that of brokers and experts.  This can be explained by the fact that garages are almost only relevant for motor claims, while our network also incorporates other LoBs.

\begin{table}
\centering
\caption{Summary statistics of nodes and edges in the sets $C$ and $P$.  The column \emph{Nodes} shows the relative frequency of a specific node type within the set of parties. Similarly, the column \emph{Edges} shows the relative frequency of edges connected to the respective node type in the set of all edges. As an edge always connects one claim and one party, the edge can be attributed to just the party, as is done in this table. \label{T:edgeSummary}}
\begin{tabular}{llm{1.3cm}m{1.3cm}m{1.3cm}m{1.3cm}m{1.2cm}m{1.3cm}}
\toprule
Set&Type&Nodes&Edges&Mean degree&Median degree&Min degree&Max degree \\ \midrule
C&Claim&-&-&3.79&3&1&42\\ \midrule
\multirow{4}{*}{P}&Policyholder&0.9636&0.4908 &2.06&1&1&20274\\
&Broker&0.0039&0.2782 & 261.00&42&1&19830\\
&Expert&0.0037&0.1513&148.00&1&1&125951\\
&Garage&0.0288&0.0797&10.10&1&1&10436\\ \bottomrule
\end{tabular}
\end{table}

\subsection{Fraudulent structures}\label{sec:fraudstruc}
A cycle in a network is a set of nodes connected via edges, where all nodes and edges are distinct and the last node is connected to the first node \citep{newman2010networks}.
In Figure \ref{simpleNetwork}, C1, P2, C3 and P3 form a four node cycle (blue edges), also called a 4-cycle.
In addition, since claims C2 and C5 are both linked to party P4 there is a cycle of six nodes, made up of C1, P1, C2, P4, C5 and P3 (orange edges).  We refer to these cycles as 6-cycles.
These cyclic structures are examples of incidents involving multiple policyholders who file claims together.
\begin{figure}
\centering
\includegraphics[scale=1]{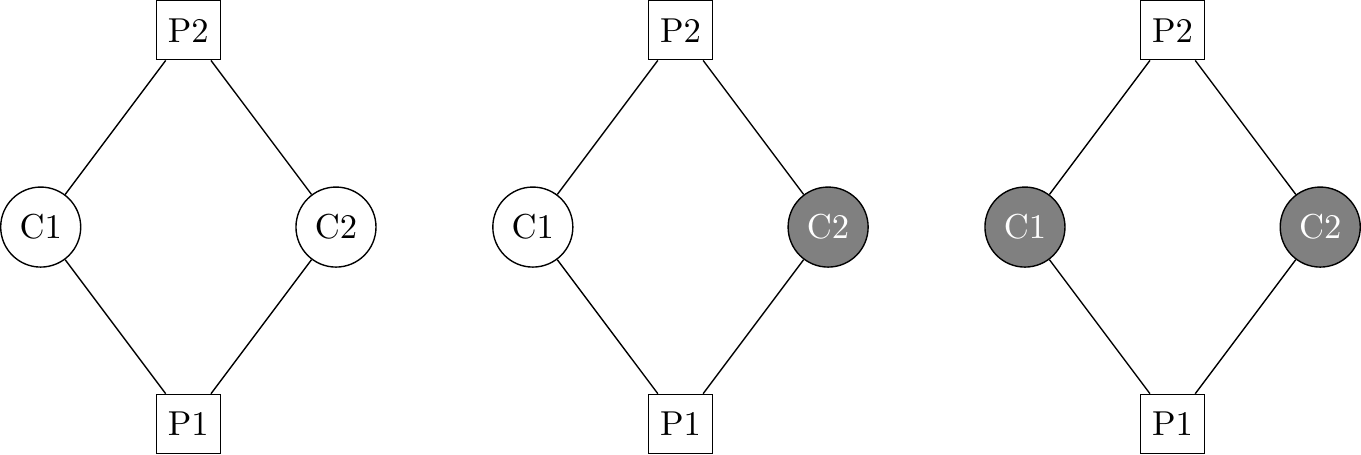}
\caption{4-cycles with zero (left), one (middle) or two (right) fraudulent claims. Fraudulent claims are colored dark gray and non-fraudulent claims are white.\label{fig:diamonds} }
\end{figure}

The most simple structure in a bipartite network is a 4-cycle, as seen in Figure \ref{fig:diamonds} where white and dark gray circles represent non-fraudulent and fraudulent claims, respectively.
In these structures zero, one or both claims can be fraudulent.
Table \ref{T:diamonds} shows the relative frequency of each type of 4-cycle among all 4-cycles with known labels in our network.
For 4-cycles where both party nodes are people (policyholder, broker or expert), over $70\%$ have two fraudulent claims, $16.33\%$ of 4-cycles have two legitimate claims and $12.24\%$ have one claim with each label.
Including 4-cycles where one of the parties can be a company, $34.87\%$ and $35.66\%$ of all 4-cycles have two fraudulent and two non-fraudulent nodes, respectively.
In contrast, with a relative frequency of $29.47\%$, the ratio of 4-cycles with one claim of each label is lower.
As such, in the 4-cycles with all labelled claims, fraudulent as well as non-fraudulent claims group together to some extent.
\begin{table}
\centering
\caption{Relative frequency of 4-cycles with zero, one and two fraudulent claims among 4-cycles with known labels.\label{T:diamonds} }
\begin{tabular}{llrp{3cm}r}
\toprule
&&\multicolumn{3}{c}{Party nodes}\\
&&Two people&One person \& one company & Total\\ \midrule
\multirow{3}{3cm}{Number of fraudulent claim nodes}&Zero &$16.33\%$&$36.56\%$&$35.66\%$\\
&One&$12.24\%$&$30.29\%$&$29.47\%$\\
&Both&$71.43\%$&$33.15\%$&$34.87\%$\\ \bottomrule
\end{tabular}
\end{table}

\begin{figure}
\centering
\includegraphics[scale=0.68]{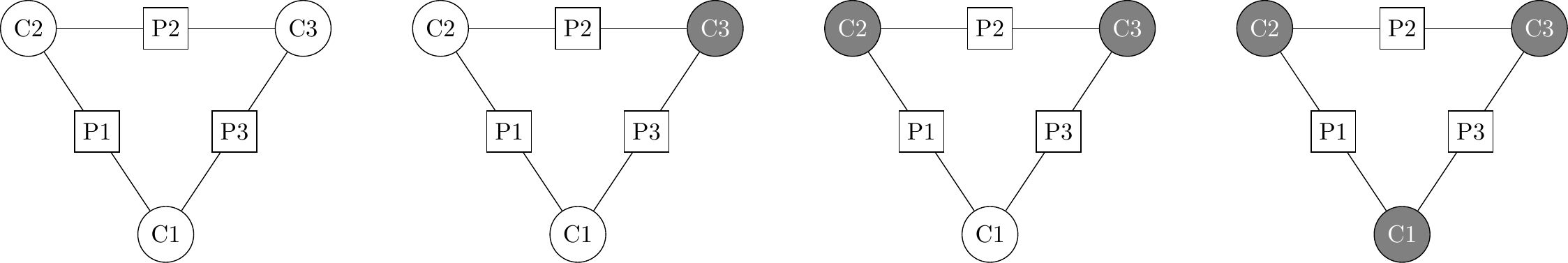}
\caption{6-cycles with zero, one, two and three fraudulent claims (from left to right). Fraudulent claims are colored dark gray and non-fradulent claims are white.}\label{triangles}
\end{figure}

6-cycles can have zero, one, two or three fraudulent claims, as seen in Figure \ref{triangles}.
We extract from the network all 6-cycles where all the claims have a known label and look at the distribution of fraudulent and non-fraudulent labels in these 6-cycles.
When a labelled claim is  part of such a 6-cycle, fraudulent claims appear on average in 1.6 6-cycles whereas non-fraudulent claims are on average in 1.4 6-cycles.
Among all 6-cycles with at least one fraudulent claim, the relative frequency of 6-cycles with
one, two and three fraudulent claims is $42.13\%$, $46.88\%$ and $10.69\%$, respectively.
Similarly, among all 6-cycles with at least one non-fraudulent claim, the relative frequency of 6-cycles with zero, one and two fraudulent claims is equal to $20.86\%$, $61.64\%$ and $17.36\%$, respectively.
As such, both fraudulent and non-fraudulent claims tend to group together in 6-cycles.

Table \ref{T:churncounts} shows the relative frequency of fraudulent and non-fraudulent claims in the second and fourth order neighborhoods of claims with a known label.
Hereto, we calculate for each labelled claim the ratio of fraudulent and non-fraudulent claims relative to the size of the respective neighborhood, that is, including the unknown claims as well.
In the second order neighborhood of non-fraudulent claims the ratio of non-fraudulent claims is $0.394\%$, on average, and the ratio of fraudulent claims is $0.154\%$, on average. 
In contrast, for fraudulent claims, the ratio of non-fraudulent claims is $0.223\%$, an average, and the ratio of fraudulent claims is $0.269\%$, on average, in the second order neighborhood.
More fraudulent claims than non-fraudulent claims appear in the neighborhood of fraudulent claims and vice versa for the neighborhood of non-fraudulent claims.
In the fourth order neighborhood, this also holds for non-fraudulent claims, but not for the fraudulent ones.

\begin{table}
\centering
\caption{Relative frequency of fraudulent and non-fraudulent claims among all claims in the second and fourth order neighborhoods of labelled claims.}\label{T:churncounts}
\begin{tabular}{llcccc}
\toprule
&&\multicolumn{2}{c}{Relative frequency}\\
Neighborhood&Label&  Fraudulent claims& Non-fraudulent claims\\ \midrule
\multirow{2}{*}{Second order}&Non-fraudulent&$0.154\%$&$0.394\%$\\
&Fraudulent&$0.269\%$&$0.223\%$\\
\multirow{2}{*}{Fourth order}&Non-fraudulent&$0.0884\%$&$0.196\%$\\
&Fraudulent&$0.0894\%$&$0.135\%$\\ \bottomrule
\end{tabular}
\end{table}
We find some empirical evidence of structural similarities in the network among fraudulent claims on the one hand and among non-fraudulent claims on the other hand.
Claims with the same label tend to be more connected to each other.
In addition, the claims with opposite labels are less connected to each other.
This establishes some evidence for homophily, especially among the fraudulent claims \citep{park2007distribution}.

\begin{figure}
\centering
\includegraphics[scale=0.9]{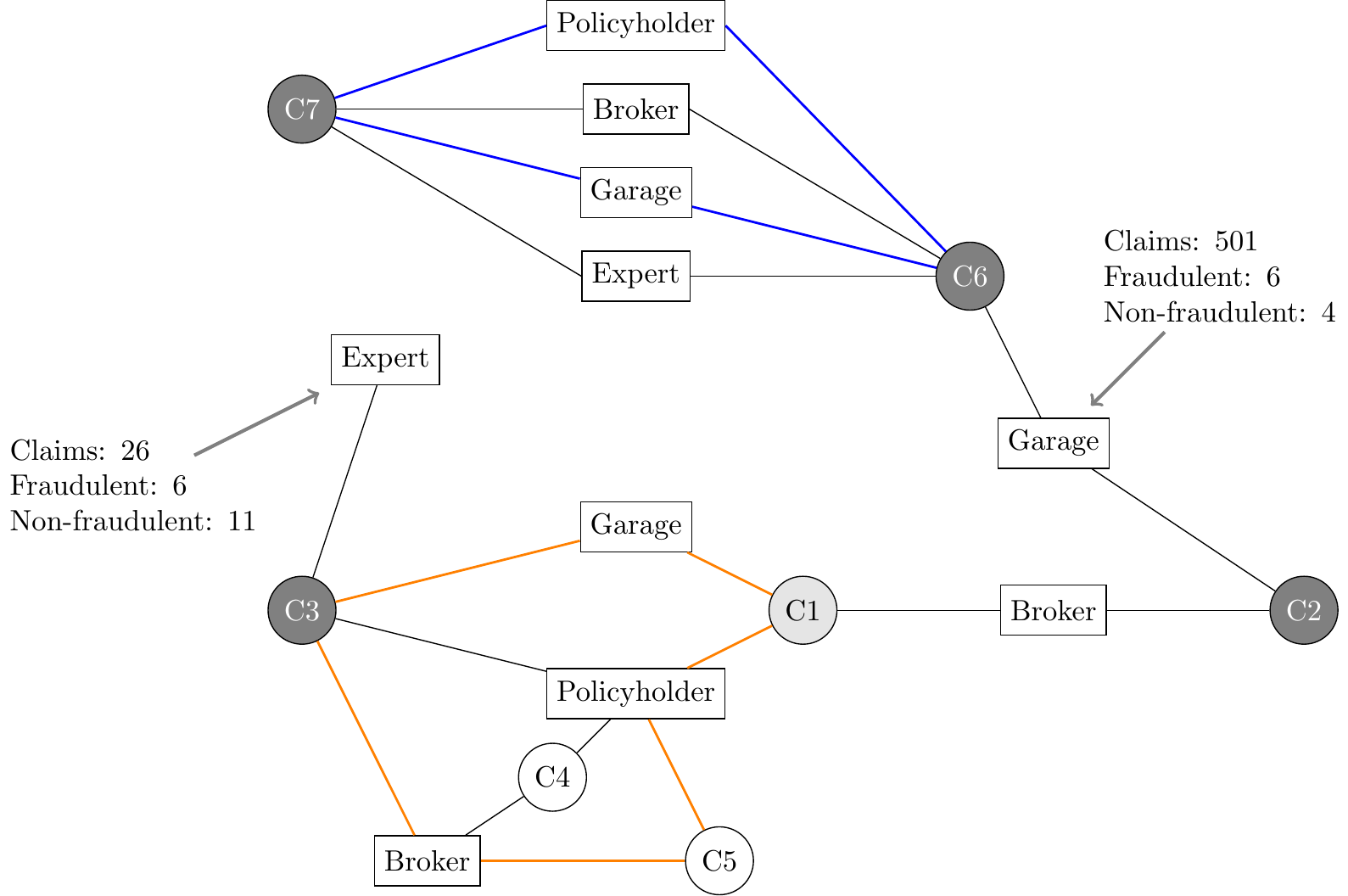}
\caption{Excerpt from the real network, illustrating the network's complexity. In the network, the dark gray, light gray and white circles represent fraudulent, unknown and non-fraudulent claims, respectively. The nodes connected with blue edges form a 4-cycle and the nodes connected with orange edges form a 6-cycle. \label{fig:exampleNetwork} }
\end{figure}

However, finding new (yet undetected) fraudulent claims through exploration of the network alone is a complex task as Figure \ref{fig:exampleNetwork} illustrates.
This excerpt from the real network shows ten parties and seven claims of which four are fraudulent (dark gray), two are non-fraudulent (white) and one is unknown (light gray).
The expert on the left side is involved in 26 claims where six are known fraudulent and eleven are known non-fraudulent.
The garage on the right has 501 claims where six are labelled as fraudulent and four are labelled as non-fraudulent.
Let C1 be a new claim that the claim handlers need to screen to decide whether it is suspicious or not.
From the intrinsic features in Table \ref{T:dataset} they can easily see that the policyholder has an older fraudulent claim.
However, detecting the influence of other nodes in the network on the claim's fraud propensity is not obvious, and requires taking the network structure into account.
For example, in this excerpt, C1 is only three edges away from the expert on the left who has a number of fraudulent claims.
Moreover, the broker on the right is part of the first order neighborhood of C1 while being connected to a proven fraudulent claim involving a garage with a dubious reputation.
Claim C1 forms a 4-cycle with the fraudulent claim C3 and a 6-cycle with fraudulent claim C3 and the non-fraudulent claim C5.
Similarly, claim C2 is connected to a garage involved in 501 claims.
It would not be feasible for the claim handlers to look at all those claims to establish connections to known fraudulent claims and to unravel structures such as the 4-cycle that exists between claims C6 and C7.

This example illustrates the complexity of manually exploring the network to find fraudulent claims and motivates the need for a more automated approach.

\section{Supervised learning with intrinsic and network derived features}\label{sec:frauddetectionmodel}
We develop an analytical model for flagging suspicious claims inspired by our empirical findings of homophily among the fraudulent claims in the network.
In Section \ref{subsec:birank}, we use a ranking algorithm based on the personalized PageRank of \cite{page1999pagerank} to score the nodes in the network with respect to their exposure to known fraudulent claims.  
We then extract features from the network in Section \ref{sec:networkfeatures} and combine them with the intrinsic features listed in Table \ref{T:dataset} in a predictive model to identify the most suspicious claims in Section \ref{subsec:analyticmodel}.

\subsection{Computing fraud scores}\label{subsec:birank}
The key principle of PageRank is to score ---and as such to rank--- a node in a network based on both the number of connecting nodes as well as the rank of the nodes that provide these connections.
The algorithm can be personalized to bring out nodes that are important from the perspective of a specific set of  source nodes, such as the known fraudulent claims in our problem setting. The resulting scores are then biased or personalized towards the source nodes, reflecting a preference for these nodes. 
\cite{he2017birank} formalized the (personalized) ranking of nodes in a bipartite network. We apply their BiRank algorithm to our network of claims and parties.

\subsubsection{BiRank algorithm}
The BiRank algorithm takes as inputs a network and a \emph{query vector} and calculates a score or ranking for each node in the network.
The query vector encodes some prior knowledge, such as fraud.
In what follows, we use $c_i$ and $p_j$ to denote the score of node $c_i$ and $p_j$, respectively. The collection of scores across all claim nodes is stored in $\mathbf{c}$, while $\mathbf{p}$ is the vector with the scores of the nodes of type party.

BiRank iteratively computes a node's score as a sum of the scores from the connected nodes.
Let $w_{ij}$ be  a non-negative number expressing the weight on the edge between nodes $c_i$ and $p_j$, where $w_{ij}$ is 0 if the nodes are unconnected.
Then
\[
c_i=\sum_{j=1}^{n_P}w_{ij}p_j\quad \textrm{and} \quad p_j=\sum_{i=1}^{n_C}w_{ij}c_i
\]
for claims and parties, respectively.
To ensure convergence and stability,
\cite{he2017birank} propose a symmetric normalization that smooths the edge weight by the degree of the two connected nodes simultaneously.
This gives the normalized scores
\begin{equation}\label{eq:rank}
c_i=\sum_{j=1}^{n_P}\frac{w_{ij}}{\sqrt{d_i}\sqrt{d_j}}p_j\quad \textrm{and} \quad p_j=\sum_{i=1}^{n_C}\frac{w_{ij}}{\sqrt{d_i}\sqrt{d_j}}c_i
\end{equation}
where $d_i$ (respectively $d_j$) is the degree of node $c_i$ (respectively $p_j$), as introduced in Section \ref{sec:bipartNetwork}.
The normalization dampens the contribution of high degree nodes and gives better quality results \citep{he2017birank}.

Query vectors  incorporate prior information or belief in the node ranking, and steer it towards the fraudulent claims, by factoring them directly into the ranking process.
Let $\mathbf{c^0}$ and $\mathbf{p^0}$ be query vectors of length $n_C$, respectively $n_P$, for the claims and parties.
To incorporate the query vectors into the ranking we expand \eqref{eq:rank} as
\begin{equation}\label{eq:BiRank1}
c_i=\alpha \sum_{j=1}^{n_P}\frac{w_{ij}}{\sqrt{d_i}\sqrt{d_j}}p_j + (1-\alpha)c_i^0\quad \textrm{and} \quad p_j=\beta\sum_{i=1}^{n_C}\frac{w_{ij}}{\sqrt{d_i}\sqrt{d_j}}c_i + (1-\beta)p_j^0
\end{equation}
where $\alpha\in[0,1]$ and $\beta\in[0,1]$ are parameters that control the trade-off between the importance of the network structure and the query vector.
Writing the BiRank equation \eqref{eq:BiRank1} in matrix form gives
\begin{equation}
\label{eq:BiRank2}
\mathbf{c}=\alpha \boldsymbol{S} \mathbf{p} +(1-\alpha)\mathbf{c}^0\quad \textrm{and} \quad \mathbf{p}=\beta \boldsymbol{S}^T \mathbf{c} +(1-\beta)\mathbf{p}^0,
\end{equation}
where $\boldsymbol{S}=\boldsymbol{D}_C^{-\frac{1}{2}}\boldsymbol{W} \boldsymbol{D}_P^{-\frac{1}{2}}$ is the symmetrically normalized weight matrix.
When using the BiRank algorithm to compute fraud scores, the ranking vectors $\mathbf{c}$ and $\mathbf{p}$ are first randomly initialized and then computed iteratively until convergence to a unique stationary point, which is theoretically guaranteed as outlined in \citep{he2017birank}.

\subsubsection{Fraud specialized query vector}
In our insurance context, the goal is to rank the nodes relative to fraud using the information about known fraudulent claims to guide the process.
The strategy is inspired by \cite{van2016gotcha} who used traditional personalized PageRank (instead of BiRank) to propagate fraud through the network.
Most of the claim nodes in the network have an unknown label, but we use the known fraudulent claims as prior information in the query vector.
However, only claims can be fraudulent, not parties.
Therefore, we identify prior information for the nodes in $C$ via $\mathbf{c}^0$, while $\mathbf{p}^0\equiv \mathbf{0}$.
We adjust \eqref{eq:BiRank2} accordingly and set $\beta=1$, since only the network structure matters in the absence of the query vector.
This gives
\[
\mathbf{c}=\alpha S \mathbf{p} +(1-\alpha)\mathbf{c}^0\quad \textrm{and} \quad \mathbf{p}= S^T \mathbf{c}.
\]
Algorithm \ref{alg1} summarizes the update rules for the iterative BiRank algorithm.
These rules are repeated until either the relative change in scores between two iterations is below a certain threshold or a maximum number of iterations is reached.
\begin{algorithm} 
\caption{BiRank algorithm for computing fraud scores in a network of insurance claims and parties. Adapted from Algorithm 1 in \citep{he2017birank}. We omit the query vector $\mathbf{p}^0$ and set $\beta=1$.} 
\label{alg1} 
    \KwIn{Weight matrix $\boldsymbol{W}$, query vector $\mathbf{c}^0$ and hyperparameter $\alpha$;}
    \KwOut{Ranking vectors  $\mathbf{c}$ and $\mathbf{p}$;}
    Symmetrically normalize $\boldsymbol{W}$: $\boldsymbol{S}=\boldsymbol{D}_P^{-\frac{1}{2}}\boldsymbol{W} \boldsymbol{D}_C^{-\frac{1}{2}}$\;
    Randomly initialize $\mathbf{c}$ and $\mathbf{p}$\;
    \While{stopping criteria is not met}
    {
            $\mathbf{c}\leftarrow\alpha \boldsymbol{S} \mathbf{p} +(1-\alpha)\mathbf{c}^0$\;
            $\mathbf{p}\leftarrow \boldsymbol{S}^T \mathbf{c}$ \;
            }
    \Return{ $\mathbf{c}$ and $\mathbf{p}$};\
\end{algorithm}

\begin{figure}
\centering
\includegraphics[width=.8\linewidth]{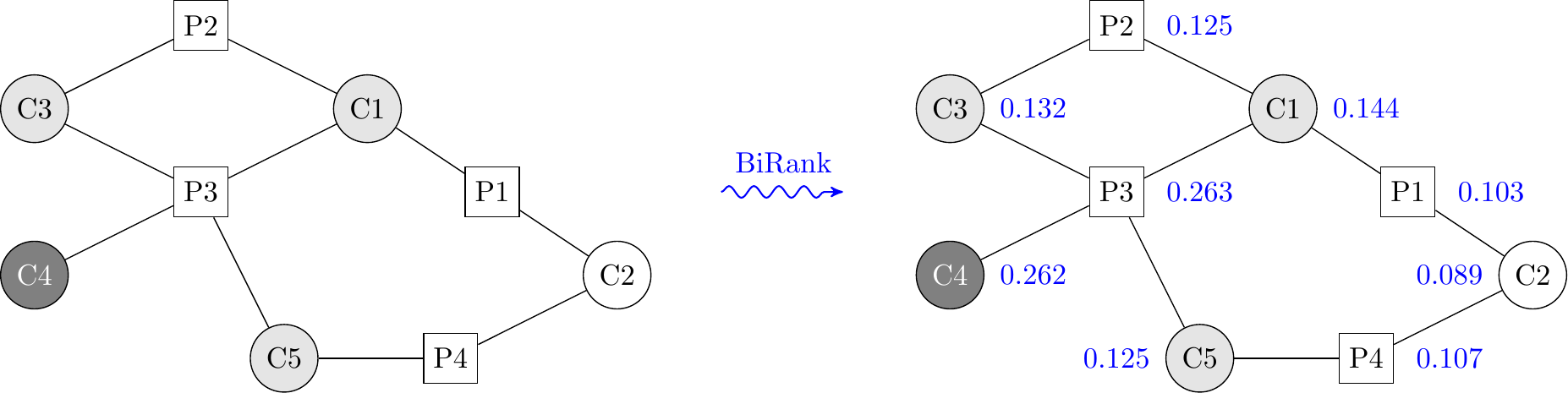}
\caption{A sample network before and after applying the BiRank algorithm to rank the nodes with respect to fraud. Claim C4 is a known fraud and the source of fraud influence. Node C2 is known non-fraud. The network is unweighted and $\alpha=0.85$. The fraud scores are shown in blue.}
\label{fig:featurisation}
\end{figure}

Figure~\ref{fig:featurisation} revisits the sample network in Figure~\ref{simpleNetwork} and shows the network before and after applying the BiRank algorithm with a query vector for fraud.
Claim C4 is known to be fraudulent and claim C2 is known to be non-fraudulent. The other claims have an unknown label. The query vector specifies claim C1 as fraudulent, i.e., $c^0_4=1$, while the other claims do not contribute any information in the query vector, i.e., $c^0_1=c^0_2=c^0_3=c^0_5=0$.
We set $\alpha=0.85$ in Algorithm~\ref{alg1}.
 After running the BiRank algorithm on this unweighted network, all nodes obtain a fraud score, indicated by the  numbers in blue.

\subsection{Network featurization}\label{sec:networkfeatures}

Based on the network structure, the labels and the computed fraud scores, we can extract several features from the network.
This process is known as network featurization.
We distinguish score features and neighborhood features as motivated in Section \ref{sec:scorefeatures} and \ref{sec:nbhfeatures}, respectively.
We refer to the set of extracted features as \emph{network features}.

\subsubsection{Features based on fraud score}\label{sec:scorefeatures}
\begin{table}
\centering
\caption{\label{T:scoreFeatures} Features extracted from the network structure, the labels and the fraud scores.}
\scalebox{0.8}{\begin{tabular}{lcp{13cm}}
\toprule
Name&Order&Description\\
\midrule
 \texttt{scores0}&0&The node's fraud score\\
 \texttt{n1.q1}&1&The first quartile of the empirical distribution of the fraud scores in the node's first order neighborhood\\
 \texttt{n1.med}&1&The median of the empirical distribution of the fraud scores in the node's first order neighborhood\\
 \texttt{n1.max}&1&The maximum of the empirical distribution of the fraud scores in the node's first order neighborhood\\
 \texttt{n2.q1}&2&The first quartile of the empirical distribution of the fraud scores in the node's second order neighborhood\\
 \texttt{n2.med}&2&The median of the empirical distribution of the fraud scores in the node's second order neighborhood\\
 \texttt{n2.max}&2&The maximum of the empirical distribution of the fraud scores in the node's second order neighborhood\\ \bottomrule
\end{tabular}}
\end{table}

Inspired by \cite{van2016gotcha} we compute network features from the fraud scores in the zeroth, first and second order neighborhoods of each node.
Table~\ref{T:scoreFeatures} shows the list of extracted \emph{score features}.
The zeroth order neighborhood of a node is the node itself and the feature we extract is the node's fraud score.
For the first and second order neighborhood features, we look at the distribution of the fraud scores in the respective neighborhoods of the claim and compute three features in each case.
These are the scores' first quartile, the median and the maximum, capturing the variation in the exposure to fraud.
For example, a high maximum indicates close proximity to at least one fraudulent node whereas a high first quartile indicates proximity to several fraudulent nodes.
The first order neighborhood typically consists of only a few nodes whereas the second order neighborhood is much larger and thus provides more variation in scores and the resulting features.
For example, returning to the sample network in Figure~\ref{fig:featurisation}, claim C1 has the score feature values shown in Table~\ref{T:scoreFeaturesSample}.
\begin{table}
\centering
\caption{Score feature values for claim C1 in the sample network in Figure \ref{fig:featurisation}.}\label{T:scoreFeaturesSample}
\scalebox{0.8}{\begin{tabular}{lcl}
\toprule
Name&Order&Value\\
\midrule
 \texttt{scores0}&0&0.1440\\
 \texttt{n1.q1}&1&0.1140\\
 \texttt{n1.med}&1&0.1250\\
 \texttt{n1.max}&1&0.2630\\
 \texttt{n2.q1}&2&0.1160\\
 \texttt{n2.med}&2&0.1285\\
 \texttt{n2.max}&2&0.2620\\ \bottomrule
\end{tabular}}
\end{table}

\subsubsection{Features based on neighborhood}\label{sec:nbhfeatures}
\begin{table}
\centering
\caption{\label{T:nbhFeatures} Features extracted from the neighborhood structure and the labels.}
\scalebox{0.8}{\begin{tabular}{lcp{13cm}}
\toprule
Name&Order&Description\\
\midrule
 \texttt{n1.size}&1&The number of nodes in the node's first order neighborhood\\
 \texttt{n2.size}&2&The number of nodes in the node's second order neighborhood\\
 \texttt{n2.ratioFraud}&2&The number of known fraudulent claims in the node's second order neighborhood divided by  \texttt{n2.size}\\
 \texttt{n2.ratioNonFraud}&2&The number of known non-fraudulent claims in the node's second order neighborhood divided by  \texttt{n2.size}\\
 \texttt{n2.binFraud}&2&A binary value indicating whether there is a known fraudulent claim in the node's second order neighborhood\\ \bottomrule
\end{tabular}}
\end{table}
Table~\ref{T:nbhFeatures} lists the extracted \emph{neighborhood features}.
First, we look at the number of nodes in a node's neighborhood.
The size of the first order neighborhood is the number of parties that are involved in the claim while the size of the second order neighborhood is the number of claims in which these involved parties are engaged in.
Second, following the framework of \cite{lu2003link} we compute three so-called link-based features using the labels of claims in a node's second order neighborhood.
These are the ratio of known fraudulent as well as non-fraudulent claims, versus the size of the neighborhood. These represent the homophilic nature of fraud.
The third feature indicates whether one of the involved parties is linked to a fraudulent claim, i.e., whether there is a fraudulent claim in its second order neighborhood.
This feature is one of the business rules used by the insurance company to flag suspicious claims for further investigation.
The values of the extracted neighborhood features for claim C1 in Figure \ref{fig:featurisation} are listed in Table~\ref{T:nbhFeaturesValues}.
\begin{table}
\centering
\caption{ Neighborhood feature values for claim C1 in the sample network in Figure \ref{fig:featurisation}.\label{T:nbhFeaturesValues}}
\scalebox{0.8}{\begin{tabular}{lcl}
\toprule
Name&Order&Value\\
\midrule
 \texttt{n1.size}&1&3\\
 \texttt{n2.size}&2&4\\
 \texttt{n2.ratioFraud}&2&0.25\\
 \texttt{n2.ratioNonFraud}&2&0.25\\
 \texttt{n2.binFraud}&2&1\\ \bottomrule
\end{tabular}}
\end{table}

\subsection{Fraud analytical model}\label{subsec:analyticmodel}
We build a supervised learning model to detect suspicious claims.
The model can use features from three groups; the intrinsic features described in Section \ref{sec:localfeatures}, the score features and the neighborhood features described in Section \ref{sec:networkfeatures}.
The features in each group are denoted with $x^{\texttt{intr}}$, $x^{\texttt{score}}$ and $x^{\texttt{nbh}}$, respectively.


Three different labels are present in the data, namely fraud, non-fraud and unknown.
We adopt a one-vs.-all classification strategy by transforming the label to a binary target feature \citep{bishop2006pattern}.
We do this in two ways.
First, we build a model that distinguishes claims with a known label from those with an unknown label.
Hereto, we create the target feature $y^{\texttt{known}}$ by setting
\begin{equation*}
y_i^{\texttt{known}}= \left\{ \begin{array}{ll}
1& \Leftrightarrow	l_i \in \{\textrm{fraud, non-fraud}\} \\
0& \Leftrightarrow	l_i \in \{\textrm{unknown}\}, \\
\end{array} \right.
\end{equation*}
with $l_i$ the label of claim $c_i$.
This results in the dataset $\mathcal{D}^{\texttt{known}}$ with features and target
\[
x=\{x^{\texttt{intr}}, x^{\texttt{score}}, x^{\texttt{nnbh}}\},\ \quad y=y^{\texttt{known}}.
\]
The claims where the target feature equals 1, are the claims that underwent a fraud investigation.
These claims were considered suspicious by the claim handlers, possibly because they fulfilled some of the currently implemented business rules to highlight fraud.
Analysis of this dataset will give insights into how claims are selected for further investigation at the insurance company. 
The findings of a supervised learning model for $y^{\texttt{known}}$ may also inspire the creation of new, additional business rules.

Second, we build a model that distinguishes claims with a known \emph{fraud} label from the rest of the claims. We create the target feature $y^{\texttt{fraud}}$ by setting
\begin{equation*}
y_i^{\texttt{fraud}}= \left\{ \begin{array}{ll}
1& \Leftrightarrow	l_i \in \{\textrm{fraud}\} \\
0& \Leftrightarrow	l_i \in \{\textrm{non-fraud, unknown}\}. \\
\end{array} \right.
\end{equation*}
This results in the dataset $\mathcal{D}^{\texttt{fraud}}$ with features and target
\[
x=\{x^{\texttt{intr}}, x^{\texttt{score}}, x^{\texttt{nnbh}}\},\ \quad y=y^{\texttt{fraud}}.
\]
This dataset will help in distinguishing fraudulent claims from the rest of the claims. We use it to investigate which features are discriminative for fraud.

We analyze the data sets  $\mathcal{D}^{\texttt{known}}$ and  $\mathcal{D}^{\texttt{fraud}}$ with two supervised learning techniques, namely logistic regression and random forests.

Logistic regression is a statistical technique used to model the probability of a certain event or class, in our case whether a given claim is fraudulent.
Given a training set, logistic regression expresses the probability of success, $P(fraud)$, as follows:
\[
P(fraud)=\frac{1}{1+e^{-(b_0+\sum b_j x_j)}}
\]
where $x_j$ are the features in the data set and $b_0,b_j$ are the parameters to be estimated by the model.
We focus on logistic regression for model building because of its popularity in the insurance industry and its interpretability.
This is in line with our goal to understand which features are predictive for the occurrence of the target variable.
These discriminative characteristics can then be transferred into business rules that claim handlers can easily understand and deploy when deciding whether an incoming claim should be flagged for investigation.

In contrast, random forests are powerful ensemble models that are capable of discovering complex patterns in the data \citep{breiman2001random}.
Ensemble methods estimate multiple models instead of using just one. 
Random forests create a bag (called `forest') of decision trees during training that together predict the outcome.  
To avoid overfitting, random forests feature additional elements of randomness. Firstly, each decision tree is trained on a bootstrap copy of the training data set. Secondly, when performing splits of nodes in the decision trees, only a random subset of features is considered. 
In this way, diverse and independent base models are created.
To make a prediction, the random forest outputs the class that the majority of the trees predicted.
This results in reduced generalization error of the prediction of the final model.
Random forests are black box models, because it is not obvious how a prediction is made. However, by investigating which features tend to give informative splits in the decision trees, it is possible to estimate their importance with respect to predicting the target outcome.
As we have multiple and varied features, we use random forests to evaluate their importance before applying logistic regression to the most informative features.

\subsection{Model evaluation}\label{subsec:evaluation}
We select three measures to evaluate the performance of the fraud models as described below.

\paragraph{Receiver operating characteristic curve}
The area under the receiver operating characteristic curve (AUROC) is commonly used to measure the performance of binary classifiers \citep{hanley1982meaning}.
It measures the trade-off between the model's recall (or sensitivity) and specificity while varying the cut-off that determines whether an instance is classified as good or bad by the model.
It summarizes the performance in a single number between 0.5 (random model) and 1 (perfect model), where a higher value indicates a better performance.
For fraud, recall measures the ratio between fraudulent claims that are detected by the model and the total number of fraudulent claims.
Specificity, on the other hand, is the number of non-fraudulent claims that the model predicts as non-fraudulent divided by the number of non-fraudulent claims in the data.

\paragraph{Precision-Recall curve}
The area under the precision-recall curve (AUPR) is a performance measure for binary classifiers that is useful when the class distribution of the dataset is imbalanced \citep{saito2015precision}.
In contrast to AUROC, it is based on the model's precision and recall.
In fraud detection, precision measures the ratio of fraudulent claims that are detected by the model and the number of claims that the model predicts as fraudulent.
The precision-recall curve shows the trade-off between the recall and the precision and represents the model's performance independent of the cut-off.
The area under the curve, AUPR, summarizes the curve in a single number.
The AUPR of a perfect model is one whereas AUPR of a random model equals the ratio of the positive class or, in our case, the fraud ratio.
\paragraph{Top decile lift}
Lift is a performance measure that represents how much better a model is at detecting fraudulent claims than a random model \citep{lemmens2006bagging, verbeke2014social}.
The top decile lift (TDL), for example, is computed by dividing the proportion of fraudulent claims among the $10\%$ of claims with the highest predicted probability with the relative frequency of the fraudulent claims in the dataset.
A random model has TDL equal to one, and increasing TDL values indicate a better performance.
The TDL performance measure is valuable from a managerial perspective as it focuses on the most critical group of claims, i.e., those with the highest fraud probability.
It enables us to control whether the targeted segment of suspicous claims indeed contains actual fraudulent claims.

\section{Experimental set up} \label{sec:experimentalsetup}
To evaluate the performance of our proposed technique,
we set up experiments reflecting the deployment of the model in practice. 
Our goal is to compare the intrinsic, score and neighborhood features in their ability to detect fraud on the one hand and labelled claims on the other hand.
We first apply the BiRank algorithm to calculate fraud scores, then we extract network features and finally we build predictive models using intrinsic and neighborhood features as well as score features.

Our model evaluation procedure requires careful design since it has to reflect the use and limitations of the predictive model in a practical setting. For example, when calculating the fraud scores the fraudulent claims that serve as the source of information in the query vector $\mathbf{c}^0$, will by design obtain a high score when the BiRank algorithm is applied.
Therefore, they can not be included in the dataset for which we build the supervised fraud detection model.
To circumvent this problem, and to reflect how the model would be deployed in reality, 
we view the older claims as historic data and include fraud information about older fraudulent claims in the query vector $\mathbf{c}^0$. 
However, we use the newer claims as targets when building the analytical model.

\begin{figure}
\begin{subfigure}{0.5\textwidth}
\centering
\includegraphics[trim={0 51.5cm 0 0},clip,width=1\linewidth]{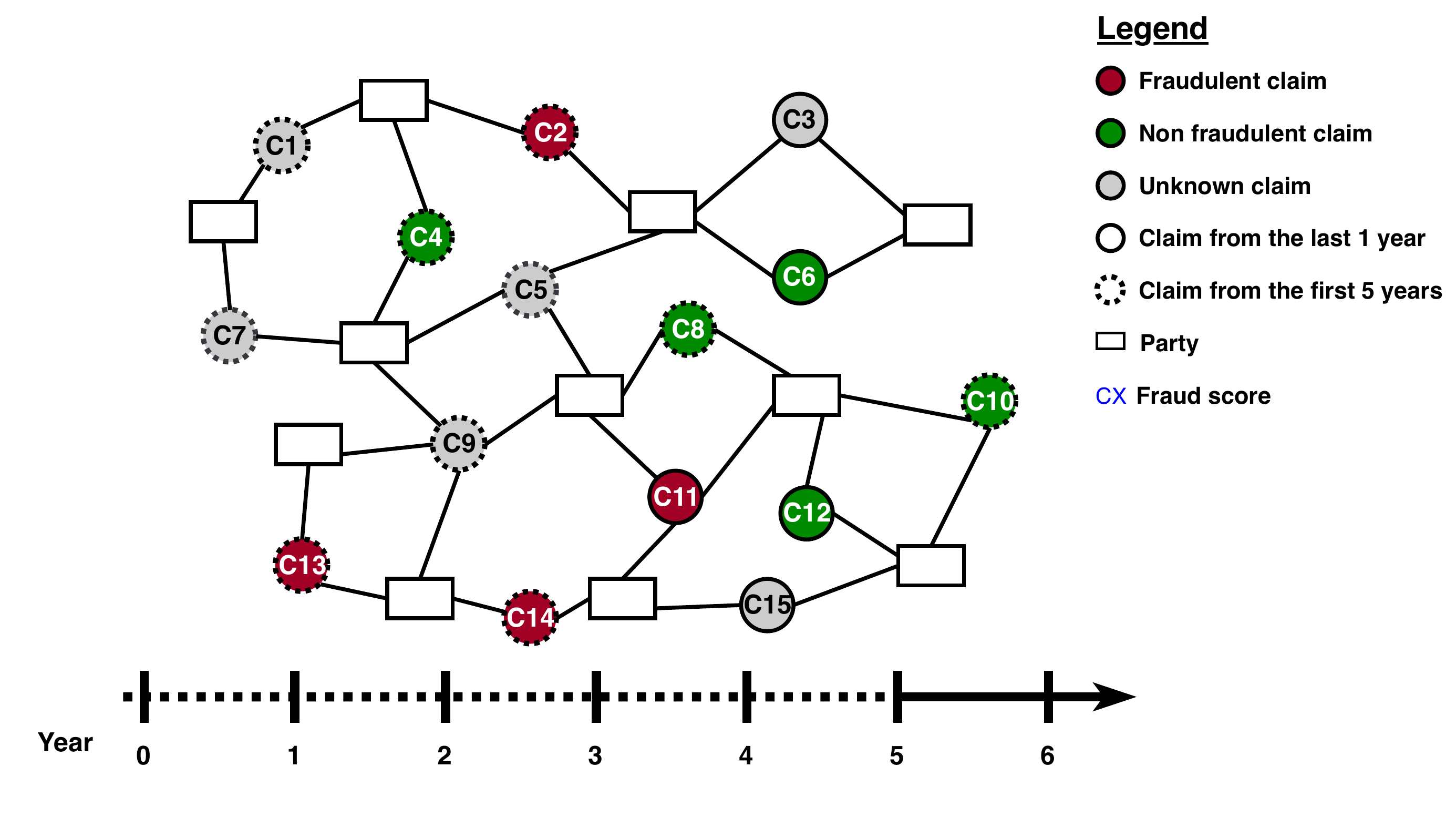}
 \caption{The whole network}
  \label{fig:expSa}
\end{subfigure}
\begin{subfigure}{0.5\textwidth}
\centering
\includegraphics[trim={0 34.4cm 0 16.9cm},clip,width=1\linewidth]{Figures/ExpSet3}
 \caption{Application of the BiRank alogrithm}
  \label{fig:expSb}
\end{subfigure}
\begin{subfigure}{0.5\textwidth}
\centering
\includegraphics[trim={0 17.3cm 0 33.8cm},clip,width=1\linewidth]{Figures/ExpSet3}
 \caption{Nodes with fraud scores}
  \label{fig:expSc}
\end{subfigure}
\begin{subfigure}{0.5\textwidth}
\centering
\includegraphics[trim={0 0 0 51.2cm},clip,width=1\linewidth]{Figures/ExpSet3}
 \caption{The most recent nodes}
  \label{fig:expSd}
\end{subfigure}
\caption{The set up for the BiRank algorithm and feature extraction. Red, green and gray circles represent fraudulent, non-fraudulent and unknown claims respectively. Rectangles represent parties. Whole and dashed lines around claims represent older and newer claims. Fraud scores are in blue. \label{fig:experimentalSetup} }
\end{figure}
\paragraph{Computation of fraud scores and network featurization}
The procedure to compute the fraud scores is shown in Figure \ref{fig:experimentalSetup}.
The claims in the data span a period of six years and we use all these claims and the involved parties to build the network. This is shown in Figure \ref{fig:expSa}.
We use the first five years of the observation period as historic knowledge about fraud and the last, most recent, observation year for model building and evaluation.
Red, green and gray nodes denote claims with the labels fraud, non-fraud and unknown, respectively.  The rectangles denote the parties.
A dashed line around a node means it was filed in the first five years and a whole line indicates the claim was filed in the most recent year.

We apply the BiRank algorithm, where the fraud specialised query vector $\mathbf{c}^0$ contains information about the known fraudulent claims of the first five years only.
Figure \ref{fig:expSb} shows how the fraud influence originates from the red fraud nodes with dashed border and spreads through the network as the red arrows indicate.
After the algorithm is applied, all nodes obtain a score. This is indicated in Figure~\ref{fig:expSc}.

Our analytical model uses only the claims that were filed in the most recent year. Therefore, we extract score and neighborhood features as described in Section~\ref{sec:networkfeatures} for these claims only, see Figure~\ref{fig:expSd}, and use these in our model building and evaluation.

\paragraph{Analytical model}
The datasets $\mathcal{D}^{\texttt{known}}$ and $\mathcal{D}^{\texttt{fraud}}$ are created by combining all labelled claims with a random sample of about 20 thousand claims that were filed in the most recent year.
We start by setting aside a test set from both $\mathcal{D}^{\texttt{known}}$ and $\mathcal{D}^{\texttt{fraud}}$.
They contain a random sample with $30\%$ of the observations in each dataset while maintaining the original class balance.  These datasets are used when measuring the out-of-sample performance of the final models. 
The remaining $70\%$ of the observations make up the training datasets.

The training datasets are highly imbalanced, with a $4.9\%$ and $1.8\%$ minority class rate in $\mathcal{D}^{\texttt{known}}$ and $\mathcal{D}^{\texttt{fraud}}$, respectively.
We use the SMOTE sampling technique to over sample the minority class and under sample the majority class \citep{chawla2002smote}.
As such, the ratio of the minority class in each sampled dataset is increased to $15\%$.
We use these sampled datasets to evaluate the features' importance using random forests.
We first tune the hyper-parameters of the random forests using ten-fold cross-validation on the training sets.
We use grid search to find the optimal parameter values for the number of trees in the forest and the number of features sampled in each split.
The search spaces for these two parameters are $\{100,300,500,700,900\}$ trees and $\{1,3,\dots, \textrm{NoF}\}$ features, where NoF is the number of features being considered. Using the optimal parameter values, we build random forest models for each of the two datasets and each feature group $x^{\texttt{intr}}$, $x^{\texttt{score}}$ and $x^{\texttt{nbh}}$. 
We then assess the importance of each feature in the constructed models.
Subsequently, we build logistic regression models where we add one feature at a time based on the features' importance, starting with the most important one.
To evaluate the performance of these logistic regression models we again use ten-fold cross-validation on the training sets. We inspect the performance increase when stepwise adding features. Finally, we build logistic regression models on the complete training set with the best performing features and evaluate the performance on the test datasets.

\section{Results}\label{sec:results}

\subsection{Performance of features}
\begin{figure}
\centering
\includegraphics[scale=0.2]{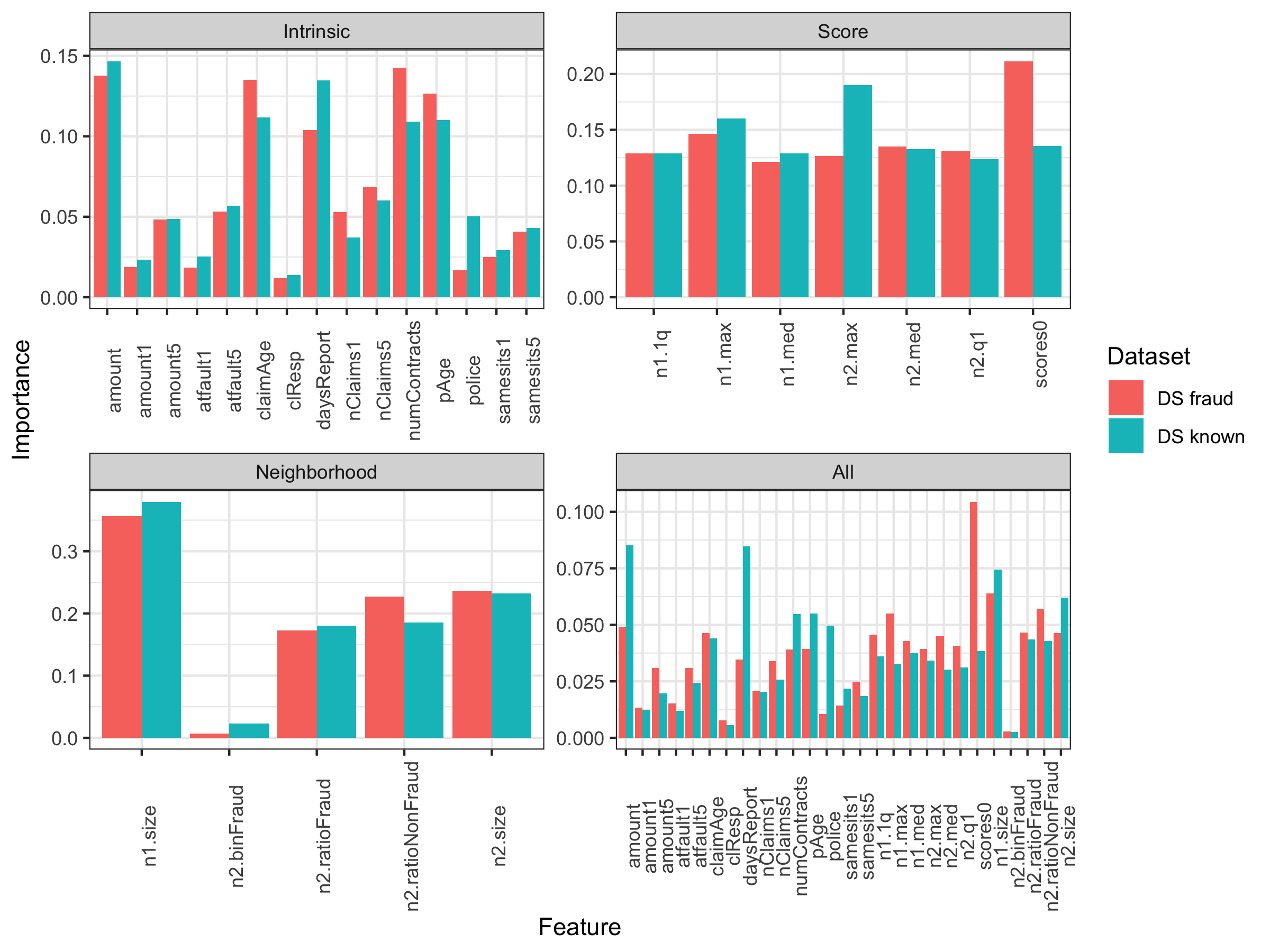}
\caption{\label{fig:importance}The importance of features in the random forest models using the two datasets ($\mathcal{D}^{\texttt{known}}$ and $\mathcal{D}^{\texttt{fraud}}$) and the four groups of features ($x^{\texttt{intr}}$, $x^{\texttt{score}}$ and $x^{\texttt{nbh}}$).}
\end{figure}
The four plots in Figure \ref{fig:importance} show the feature importance for each group of features ($x^{\texttt{intr}}$, $x^{\texttt{score}}$ and $x^{\texttt{nbh}}$) and each dataset ($\mathcal{D}^{\texttt{known}}$ and $\mathcal{D}^{\texttt{fraud}}$).
We normalize the feature importance values such that they sum to one to give a clear indication of the relative contribution of each feature.
The red bars denote the feature importance for the dataset $\mathcal{D}^{\texttt{fraud}}$ and the green bars refer to the dataset $\mathcal{D}^{\texttt{known}}$.

We detect some clear differences between the two datasets when using only intrinsic features.
The features \texttt{amount}, \texttt{daysReport} and \texttt{police} are relatively more important in the dataset $\mathcal{D}^{\texttt{known}}$, whereas the features \texttt{claimAge}, \texttt{numContracts} and \texttt{pAge} are more important in $\mathcal{D}^{\texttt{fraud}}$.
The most important score features for the dataset $\mathcal{D}^{\texttt{known}}$ are the maximum scores in the first and second order neighborhood, whereas the fraud score is most important for $\mathcal{D}^{\texttt{fraud}}$.
For the neighborhood features, the difference in feature importance between the two datasets is less obvious.
However, we observe clear differences in the importance of features when they are combined in the random forest models.
The most important features for the dataset $\mathcal{D}^{\texttt{known}}$ are intrinsic features, e.g., \texttt{amount}, \texttt{daysReport}, \texttt{numContracts}, \texttt{pAge} and \texttt{police}, whereas the most important features for $\mathcal{D}^{\texttt{fraud}}$ are score and neighborhood features, most notably \texttt{scores0}, \texttt{n1.size} and \texttt{n1.q1}.
Furthermore, intrinsic features are overall more important in $\mathcal{D}^{\texttt{known}}$ than in $\mathcal{D}^{\texttt{fraud}}$, while score features matter more in $\mathcal{D}^{\texttt{fraud}}$ than in $\mathcal{D}^{\texttt{known}}$.

\begin{figure}
\begin{subfigure}{1\textwidth}
  \centering
  \includegraphics[width=.8\linewidth]{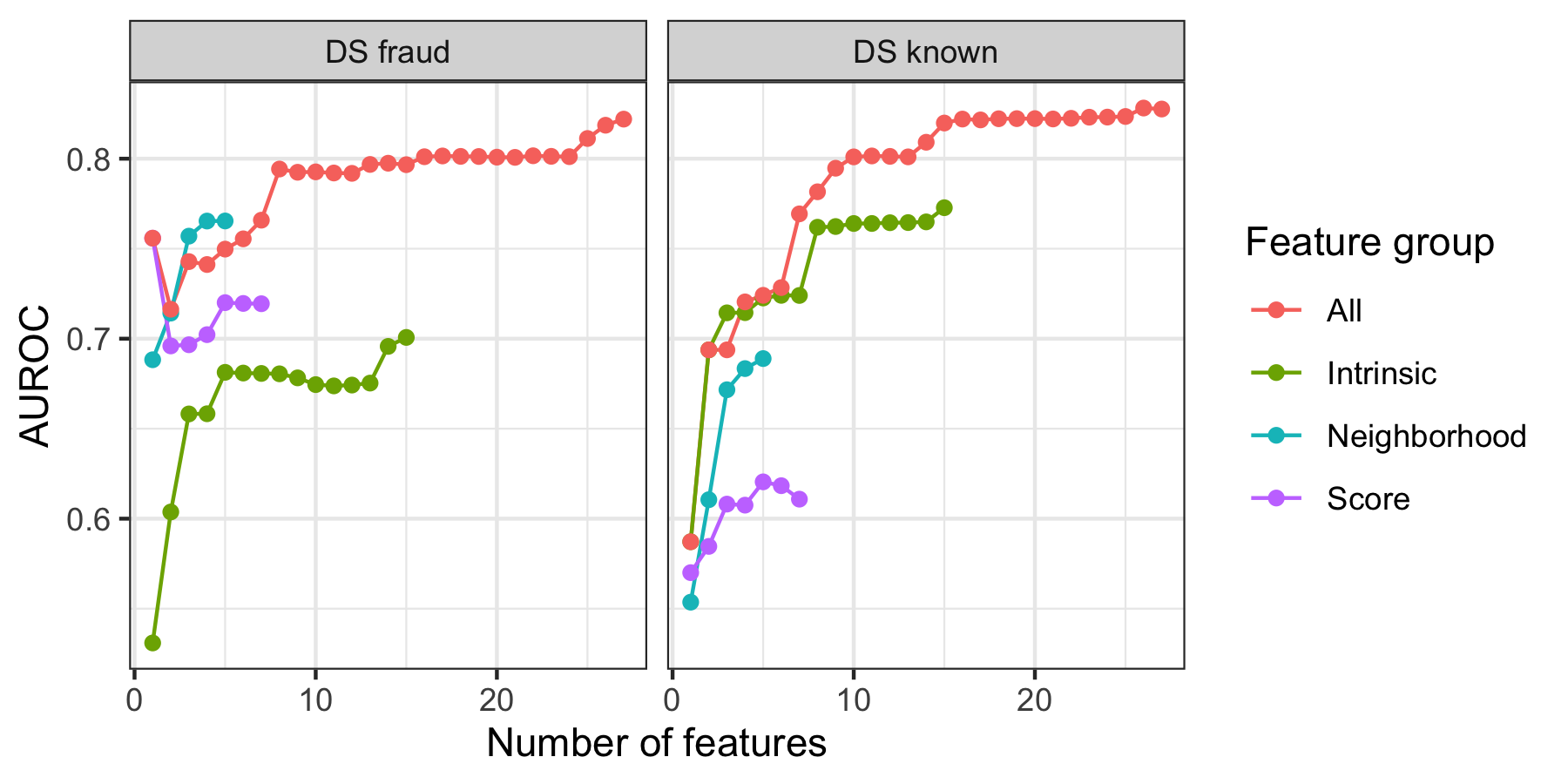}
  \caption{AUROC evaluated via ten-fold cross-validation on the training sets.}
  \label{fig:linesAUC}
\end{subfigure}
\begin{subfigure}{1\textwidth}
  \centering
  \includegraphics[width=.8\linewidth]{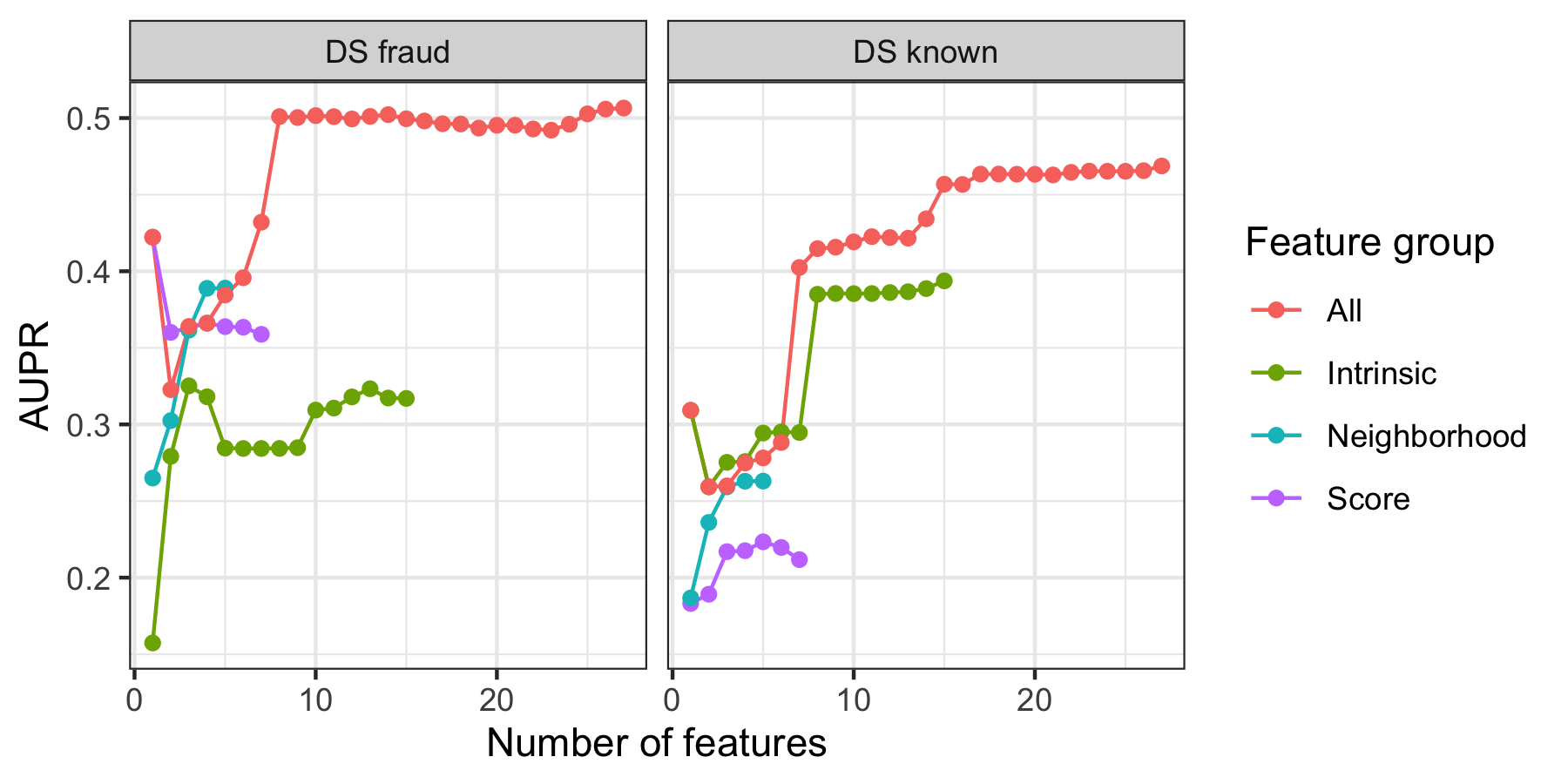}
  \caption{AUPR evaluated via ten-fold cross-validation on the training sets.}
  \label{fig:linesPR}
\end{subfigure}
\begin{subfigure}{1\textwidth}
  \centering
  \includegraphics[width=.8\linewidth]{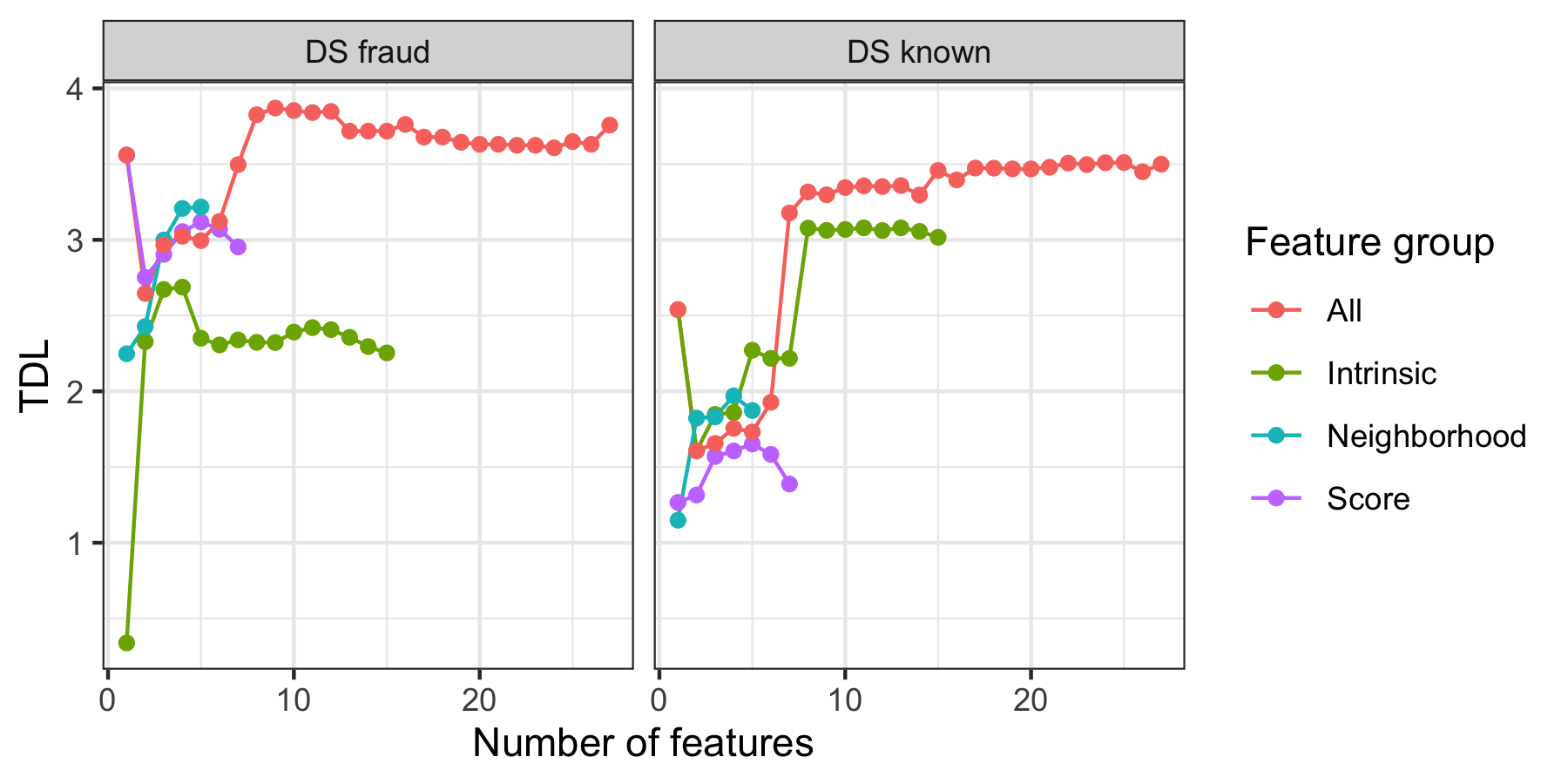}
  \caption{TDL evaluated via ten-fold cross-validation on the training sets.}
  \label{fig:linesLift}
\end{subfigure}
\caption{\label{fig:lines}The performance of logistic regression models for the two datasets ($\mathcal{D}^{\texttt{known}}$ and $\mathcal{D}^{\texttt{fraud}}$) and the four groups of features ($x^{\texttt{intr}}$, $x^{\texttt{score}}$ and $x^{\texttt{nbh}}$). We add the features in decreasing order of importance, starting with the most important feature according to the random forest models previously constructed.}
\end{figure}

Next we evaluate the prediction of the target in each dataset using logistic regression with the most important features.
We start by including only the most important feature and then add the features in decreasing order of importance according to the random forest models previously constructed. 
The plots in Figure \ref{fig:lines} show the performance measured in AUROC, AUPR and TDL evaluated via ten-fold cross-validation on the constructed training sets.
Clearly, in $\mathcal{D}^{\texttt{known}}$ the performance of the model with only intrinsic features is greater than in $\mathcal{D}^{\texttt{fraud}}$.
In contrast, the models using only score and neighborhood features perform better in $\mathcal{D}^{\texttt{fraud}}$ compared to  $\mathcal{D}^{\texttt{known}}$.

\begin{table}
\centering
\caption{Significant features in the logistic regression models with only intrinsic features.\label{T:glmLocal}}
\scalebox{0.74}{\begin{tabular}{lcccclcccc}
\toprule
& \multicolumn{4}{c}{ $\mathcal{D}^{\texttt{known}}$} & & \multicolumn{4}{c}{$\mathcal{D}^{\texttt{fraud}}$}\\
\cmidrule{2-5} \cmidrule{7-10}
 & Estimate & Std. Error & z value & Pr($>$ $|$z$|$) & & Estimate & Std. Error & z value & Pr($>$ $|$z$|$) \\
  \midrule
(Intercept) & -2.2511 & 0.0372 & -60.57 & 0.0000 & (Intercept) & -1.9177 & 0.0521 & -36.83 & 0.0000 \\
   \texttt{amount} & 0.1770 & 0.0269 & 6.59 & 0.0000 &  \texttt{numContracts}& 0.1043 & 0.0468 & 2.23 & 0.0257 \\
   \texttt{daysReport} & -0.6346 & 0.0314 & -20.24 & 0.0000 &   \texttt{amount} & 0.3627 & 0.0451 & 8.04 & 0.0000 \\
   \texttt{claimAge} & -0.3509 & 0.0328 & -10.70 & 0.0000 &  \texttt{claimAge} & -0.4830 & 0.0517 & -9.34 & 0.0000 \\
  \texttt{pAge} & -0.0443 & 0.0305 & -1.45 & 0.1461 &  \texttt{daysReport} & -0.2937 & 0.0482 & -6.09 & 0.0000 \\
  \texttt{numContracts} & 0.2422 & 0.0290 & 8.36 & 0.0000 &   \texttt{atfault5} & 0.2727 & 0.0901 & 3.03 & 0.0025 \\
   \texttt{nClaims5}& -0.1788 & 0.0860 & -2.08 & 0.0375 &   \texttt{samesits5} & -0.2994 & 0.0923 & -3.24 & 0.0012 \\
   \texttt{atfault5} & -0.2319 & 0.0945 & -2.45 & 0.0141 \\
   \texttt{police1} & 1.4167 & 0.0645 & 21.98 & 0.0000 \\
   \texttt{amount5} & 0.1805 & 0.0673 & 2.68 & 0.0074 \\
   \texttt{samesits5} & 0.1903 & 0.0664 & 2.87 & 0.0041 \\
   \bottomrule
\end{tabular}}
\end{table}

\begin{table}
\centering
\caption{Significant features in the logistic regression models with only score features. \label{T:glmScore} }
\scalebox{0.8}{\begin{tabular}{lcccclcccc}
\toprule
& \multicolumn{4}{c}{ $\mathcal{D}^{\texttt{known}}$} & & \multicolumn{4}{c}{$\mathcal{D}^{\texttt{fraud}}$}\\
\cmidrule{2-5} \cmidrule{7-10}
 & Estimate & Std. Error & z value & Pr($>$ $|$z$|$) & & Estimate & Std. Error & z value & Pr($>$ $|$z$|$)\\
  \midrule
(Intercept) & -1.7427 & 0.0276 & -63.12 & 0.0000 & (Intercept) & -1.9211 & 0.0521 & -36.86 & 0.0000 \\
  \texttt{n2.max} & 0.3075 & 0.0435 & 7.07 & 0.0000 &  \texttt{scores0} & 0.2697 & 0.0558 & 4.83 & 0.0000 \\
  \texttt{n1.max} & -0.1042 & 0.0365 & -2.85 & 0.0043 &   \texttt{n1.max} & 0.3615 & 0.0487 & 7.43 & 0.0000 \\
  \texttt{scores0} & 0.1847 & 0.0246 & 7.50 & 0.0000 &  \texttt{n2.med} & 1.2744 & 0.2926 & 4.35 & 0.0000 \\
   \texttt{n2.med} & -0.4620 & 0.1618 & -2.86 & 0.0043 &  \texttt{n2.q1} & -0.8991 & 0.2831 & -3.18 & 0.0015 \\
  \texttt{n1.1q} & -0.2348 & 0.0400 & -5.88 & 0.0000 &  \texttt{n1.1q} & -0.2336 & 0.0940 & -2.48 & 0.0130 \\
   \texttt{n2.q1} & 0.5190 & 0.1660 & 3.13 & 0.0018 &  \texttt{n1.med} & -0.1321 & 0.0769 & -1.72 & 0.0858 \\
   \bottomrule
\end{tabular}}
\end{table}

\begin{table}
\centering
\caption{Significant features in the logistic regression models with only neighborhood features. \label{T:glmNbh}}
\scalebox{0.7}{\begin{tabular}{lcccclcccc}
\toprule
& \multicolumn{4}{c}{ $\mathcal{D}^{\texttt{known}}$} & & \multicolumn{4}{c}{$\mathcal{D}^{\texttt{fraud}}$}\\
\cmidrule{2-5} \cmidrule{7-10}
 & Estimate & Std. Error & z value & Pr($>$ $|$z$|$) & & Estimate & Std. Error & z value & Pr($>$ $|$z$|$)\\
  \midrule
(Intercept) & -2.4184 & 0.0912 & -26.51 & 0.0000       &     (Intercept) & -2.1173 & 0.0587 & -36.07 & 0.0000 \\
   \texttt{n1.size} & 0.6111 & 0.0407 & 15.02 & 0.0000            &       \texttt{n1.size} & 1.1005 & 0.0740 & 14.87 & 0.0000 \\
   \texttt{n2.size} & -0.7372 & 0.0466 & -15.84 & 0.0000           &    \texttt{n2.size} & -0.6127 & 0.0726 & -8.44 & 0.0000 \\
   \texttt{n2.ratioNonFraud} & 0.3002 & 0.0281 & 10.67 & 0.0000     &     \texttt{n2.ratioNonFraud }& 0.3464 & 0.0444 & 7.80 & 0.0000 \\
   \texttt{n2.ratioFraud} & 0.0971 & 0.0281 & 3.46 & 0.0005         &     \texttt{n2.ratioFraud }& 0.3621 & 0.0495 & 7.32 & 0.0000 \\
   \texttt{n2.binFraud} & 0.7459 & 0.1017 & 7.34 & 0.0000       & \\
\bottomrule
\end{tabular}}
\end{table}

\begin{table}
\centering
\caption{Significant features in the logistic regression models with all features. \label{T:glmAll} }
\scalebox{0.7}{\begin{tabular}{lcccclcccc}
\toprule
& \multicolumn{4}{c}{ $\mathcal{D}^{\texttt{known}}$} & & \multicolumn{4}{c}{$\mathcal{D}^{\texttt{fraud}}$}\\
\cmidrule{2-5} \cmidrule{7-10}
 & Estimate & Std. Error & z value & Pr($>$ $|$z$|$) & & Estimate & Std. Error & z value & Pr($>$ $|$z$|$) \\
  \midrule
  (Intercept) & -2.5352 & 0.0436 & -58.15 & 0.0000     &     (Intercept) & -2.3559 & 0.0657 & -35.86 & 0.0000 \\
   \texttt{amount }& 0.1040 & 0.0274 & 3.80 & 0.0001             &        \texttt{n1.size} & 1.1768 & 0.0789 & 14.92 & 0.0000 \\
   \texttt{daysReport }& -0.7116 & 0.0355 & -20.05 & 0.0000      &       \texttt{n2.ratioNonFraud }& 0.2645 & 0.0446 & 5.93 & 0.0000 \\
   \texttt{n1.size} & 0.6186 & 0.0468 & 13.23 & 0.0000           &        \texttt{n1.max }& 1.2969 & 0.1129 & 11.49 & 0.0000 \\
   \texttt{n2.size} & -2.1622 & 0.0972 & -22.23 & 0.0000         &        \texttt{amount }& 0.1840 & 0.0466 & 3.95 & 0.0001 \\
   \texttt{pAge}& -0.0539 & 0.0321 & -1.68 & 0.0931             &        \texttt{n2.ratioFraud }& -0.1251 & 0.0662 & -1.89 & 0.0590 \\
   \texttt{numContracts} & 0.1874 & 0.0304 & 6.17 & 0.0000       &        \texttt{claimAge }& -0.3384 & 0.0528 & -6.41 & 0.0000 \\
   \texttt{police1 }& 1.5206 & 0.0687 & 22.13 & 0.0000           &        \texttt{n2.size }& -2.2489 & 0.1620 & -13.88 & 0.0000 \\
   \texttt{claimAge} & -0.2967 & 0.0334 & -8.89 & 0.0000         &        \texttt{n1.med }& 0.0890 & 0.0551 & 1.62 & 0.1060 \\
   \texttt{n2.ratioFraud }& -0.1833 & 0.0424 & -4.33 & 0.0000    &        \texttt{n2.max }& 0.5301 & 0.0991 & 5.35 & 0.0000 \\
   \texttt{n2.ratioNonFraud }& 0.1441 & 0.0312 & 4.62 & 0.0000   &           \\
   \texttt{n2.max} & 0.7173 & 0.0565 & 12.70 & 0.0000            &            \\
   \texttt{n1.max }& 0.9963 & 0.0727 & 13.71 & 0.0000            &            \\
\bottomrule
\end{tabular}}
\end{table}

The final models are found using stepwise logistic regression with a combination of forward and backward selection of features on the training sets.
Tables \ref{T:glmLocal}, \ref{T:glmScore}, \ref{T:glmNbh} and \ref{T:glmAll} show the results for the two datasets and the four groups of features.  
Less features are significant in the models for $\mathcal{D}^{\texttt{fraud}}$ compared to $\mathcal{D}^{\texttt{known}}$. 
The score features dominate the logistic regression model for $\mathcal{D}^{\texttt{fraud}}$ fitted on the combined set of all features.

Even though multi-collinearity of the features, in particular the extracted network features, complicates the interpretation of the sign and size of a calibrated effect, we extract some interesting findings from the output of the logistic regression models.
Notably, in the intrinsic features models, see Table \ref{T:glmLocal}, \texttt{atfault5} and \texttt{samesits5} have opposite effects in the two datasets. For example, \texttt{atfault5} has a negative coefficient in $\mathcal{D}^{\texttt{known}}$, but a positive coefficient in $\mathcal{D}^{\texttt{fraud}}$. 
Policyholders who have been at fault in multiple claims in the past years are more likely to file another fraudulent claim. 
Similarly, \texttt{samesits5} has a positive coefficient in $\mathcal{D}^{\texttt{known}}$ but a negative coefficient in $\mathcal{D}^{\texttt{fraud}}$.
In addition, the feature \texttt{police1}, which indicates that the police was called to the scene of the incident, is significant with a positive sign in $\mathcal{D}^{\texttt{known}}$ but it is not significant in $\mathcal{D}^{\texttt{fraud}}$.

Table~\ref{T:glmScore} shows the results for the models with only score features, where for example both maximum statistics are significant in $\mathcal{D}^{\texttt{known}}$. They are also the most important features according to the random forest model. In $\mathcal{D}^{\texttt{fraud}}$, the \texttt{scores0} feature is most important and has a positive coefficient, which indicates that a higher fraud score detects fraudulent claims to a greater extent.
Comparing these results to Figure~\ref{fig:lines} the \texttt{scores0} feature has a lot of predictive power on its own, as it is the first feature that is added to the model. The performance drops when more features are added, due to the multi-collinearity of the score features.

For the neighborhood features in Table~\ref{T:glmNbh}, the indicator feature \texttt{n2.binFraud} is significant for detecting the labeled claims but not for detecting fraudulent claims.

Table~\ref{T:glmAll} displays the results for the models with all features. In $\mathcal{D}^{\texttt{known}}$ twelve features are significant, with half of these being intrinsic features. Nine features are significant in $\mathcal{D}^{\texttt{fraud}}$ and only two of them are intrinsic features. 

\begin{table}
\centering
\caption{\label{T:HOperf}Performance of the four models on the test datasets.}
\begin{tabular}{lcccccc}
\toprule
&\multicolumn{3}{c}{ $\mathcal{D}^{\texttt{known}}$}&\multicolumn{3}{c}{$\mathcal{D}^{\texttt{fraud}}$}\\
\cmidrule{2-4} \cmidrule{5-7}
Features&AUROC&AUPR&TDL  &AUROC&AUPR&TDL \\ \midrule
Intrinsic& 0.691 &0.1214 & 2.85&0.662&0.0301&  2.137\\
Score& 0.634 &0.0883 & 2.25&0.660&0.0402&  2.812\\
Neighborhood& 0.681 &0.1051&  2.65&0.719&0.0481&  3.262\\
All&0.725 &0.1312 & 3.457&0.792 &0.0810 &3.824\\ \bottomrule
\end{tabular}
\end{table}

In a final step we evaluate the logistic regression models on the test datasets.
Table~\ref{T:HOperf} shows that the intrinsic model is better for dataset $\mathcal{D}^{\texttt{known}}$, while both network models perform better on $\mathcal{D}^{\texttt{fraud}}$.
Combining the different types of features available leads to the best performance on the test set, both for $\mathcal{D}^{\texttt{known}}$ as well as  $\mathcal{D}^{\texttt{fraud}}$. 

The approach is feasible for deployment as it performs well in terms of runtime.
Even though all the claims and involved parties as registered over a period of six years  were used to build the social network, running the BiRank algorithm to score the most recent claims took about ten minutes for each value of $\alpha$.  The training of the algorithms (logistic regression and random forests) is not time consuming since we work with tabular datasets with carefully engineered features (only those listed in Tables \ref{T:dataset}, \ref{T:scoreFeatures} and \ref{T:nbhFeatures}).  The extraction of score and neighborhood features was the most time consuming part. In practice, the model would be used to score new claims on a weekly or monthly basis to mark suspicious claims, in which case the number of new claims is limited and runtime reasonable.

\section{Conclusion}\label{sec:conclusion}
The insurance industry is increasingly relying on data science and machine learning to provide novel insights in its practices, such as for pricing and reserving.
Insurance companies possess a lot of data about past claims, as well as policyholders and their claim history.  In that data lies knowledge, which goes beyond simple statistics, that can be leveraged for more accurate data-driven predictions for the companies to better manage their practices and accommodate their heterogeneous customer base with tailor made policies and tariffs.  Data science applied for fraud detection is another opportunity for the companies to find hidden patterns in the data, which ultimately will help them save money and strengthen their position, by giving their customers better service and lower tariffs.

This study presents a novel approach for fraud detection in insurance using social network analytics.
We leverage an insurance company's database of claims, policyholders, brokers, experts and garages to build a bipartite network that represents the social structure between the various parties involved in claims.
Taking a holistic view across multiple lines of business, we unravel stronger ties among fraudulent claims when compared to the links between fraudulent and non-fraudulent claims.
We use this empirical finding to rank claims with respect to fraud using a personalized PageRank algorithm (in casu: BiRank with a fraud specific query vector).
The resulting fraud scores bring added value when detecting fraud and so are the score neighborhood-structure features derived from the network.
Our experiments reveal a clear difference between the features that are predictive for claims with a known label (i.e.,~investigated claims) versus claims with a confirmed fraud label.
The classical intrinsic features registered for claims and policyholders are good at distinguishing claims with a known label, which is intuitive as these are traditionally used by the insurance company to flag suspicious claims for further investigation.
Features distinguishing fraudulent claims from the other claims in the dataset are the score and neighborhood features. 
We demonstrate the fitness of the proposed network representation as well as its added value in fraud detection strategies. Not only previous fraudulent behavior of a policyholder, but also a claim's entire social network is of great importance to detect new fraudulent claims.
As such, our method is `white box', we can interpret the network derived features, as opposed to the network representation learning approaches which are currently prominent in the literature. Resent research on using features derived from variations of the personalized PageRank algorithm, has demonstrated its potential to capture complex non-linear effects and correlations between interconnected objects, such as borrowers in credit scoring \citep{oskarsdottir2019value, bravo2020evolution}.

The results of the methodology presented in this paper facilitate continued research for social network based fraud detection in insurance.
Firstly, in our methodology we omitted the time-weighting of fraud influence and of edges, an approach presented in \citet{van2016gotcha}. This would be a very interesting addition and potentially may help to better understand fraud recency in the insurance context. 
Secondly, in this paper, we used SMOTE, a state-of-the-art method that is common in the fraud literature, to rebalance the dataset. More in depth research on the difference between such methods is possible, which we would like to explore in our future work. 
Thirdly, there is selection bias with respect to the fraudulent claims, since only suspicious claims undergo a fraud investigation. Because of this potential bias, we decided to consider three types of labels: unknown, fraud and non-fraud, and deployed our one-versus-all approach with the two datasets, $\mathcal{D}^{\texttt{known}}$ and $\mathcal{D}^{\texttt{fraud}}$.
However, the availability of labelled as well as unlabelled claims is an ideal premise for a semi-supervised machine learning approach. We would like to explore such methods in order to accommodate for the low ratio of labelled claims which is an inherent problem we faced in our research. 
Such an approach was developed amongs others for rejected loan applications by \citet{li2020inferring} and could potentially help in finding the rare fraudulent claims in our case. 
Next, graph based anomaly detection methods would be an interesting approach to find the rare fraudulent claims in the network. Such methods are becoming increasingly more prominent in the network science research community. 
Finally, with the fast growing development of network representation learning, there is an opportunity to learn low dimensional vector representations for the nodes in a network. These methods use deep learning approaches to learn structural properties of the nodes in a network and the resulting representations can subsequently be used with a supervised machine learning technique to classify the nodes, like we did here for fraud.  In our future work we will especially focus on frameworks such as GraphSAGE that leverage feature rich nodes, such as the claims in our case \citep{hamilton2017inductive}.

\section*{Acknowledgement}
 This work was supported by the Ageas Research chair at KU Leuven and KU Leuven’s research council [project COMPACT C24/15/001]. This support is gratefully acknowledged.

{\small
\bibliography{references_final}}


\end{document}